\documentclass[a4paper,hidelinks, 12pt]{article}
\usepackage[utf8]{inputenc}
\usepackage{a4wide}
\usepackage{graphics}
\usepackage{amsmath}
\usepackage{amsfonts}
\usepackage{amssymb}
\usepackage{multirow}
\usepackage{subfigure}
\usepackage{cite}
\usepackage{xcolor}
\usepackage{soul}
\usepackage{mathtools}
\usepackage{hyperref}
\usepackage{enumitem}
\usepackage{threeparttable}
\usepackage[normalem]{ulem}

\renewcommand{\Re}{{\thinspace\rm Re\thinspace}}
\renewcommand{\Im}{{\thinspace\rm Im\thinspace}}

\def\dd{\text{d}}
\def\thetaW{\theta_\text{W}}
\def\lsim{\mathrel{\rlap{\raise 2.5pt \hbox{$<$}}\lower 2.5pt\hbox{$\sim$}}}
\def\gsim{\mathrel{\rlap{\raise 2.5pt \hbox{$>$}}\lower 2.5pt\hbox{$\sim$}}}

\newcommand{\half}{\textstyle\frac{1}{2}}
\allowdisplaybreaks
\numberwithin{equation}{section}
\numberwithin{table}{section}
\begin{document}
\begin{titlepage}
\begin{center}

{\large \bf {CP violation in $H^\pm \to W^\pm Z$: A physical approach for the 2HDM}}

\vskip 1cm

W. Khater,$^{a,}$\footnote{E-mail: Wkhater@birzeit.edu}
O. M. Ogreid,$^{b,}$\footnote{E-mail: omo@hvl.no}
P. Osland$^{c,}$\footnote{E-mail: Per.Osland@uib.no} and 
M. N. Rebelo$^{d,}$\footnote{E-mail: rebelo@tecnico.ulisboa.pt}

\vspace{1.0cm}

$^{a}$Department of Physics, Birzeit University, Palestine,\\
$^{b}$Western Norway University of Applied Sciences,\\ Postboks 7030, N-5020 Bergen, 
Norway, \\
$^{c}$Department of Physics and Technology, University of Bergen, \\
Postboks 7803, N-5020  Bergen, Norway, \\
$^{d}$Centro de F\'isica Te\'orica de Part\'iculas -- CFTP,\\
Instituto Superior T\'ecnico -- IST, Universidade de Lisboa, Av. Rovisco Pais, \\
P-1049-001 Lisboa, Portugal
\end{center}

\vskip 3cm

\begin{abstract}
We investigate CP violation in the process $H^\pm\to W^\pm Z$ within the framework of the two-Higgs-Doublet model (2HDM). Amplitudes are expressed transparently in terms of physical couplings. Our analysis qualitatively confirms recent results by Kanemura and Mura, for the interference between one-loop bosonic and fermionic amplitudes. Furthermore, we identify additional sources of CP violation arising from internal interference among the bosonic and also among the fermionic loop amplitudes. In the alignment limit, the asymmetry would vanish in the absence of fermionic loop contributions when the sum of the masses of the additional neutral scalars is higher than the mass of the $Z$. Our results allow for generic Yukawa couplings, and cover both explicit and spontaneous CP violation.
For inclusive $W^\pm$ and $Z$ decays, the CP violation will only manifest itself as a charge asymmetry.

\end{abstract}

\end{titlepage}

\setcounter{footnote}{0} 

\section{Introduction}

While the Two-Higgs-Doublet Model (2HDM) cannot address all shortcomings of the Standard Model (SM), one can still hope to extract useful lessons from some of its features. One open issue is whether or not there is an extended scalar sector, and, if so, does it violate CP? We shall here address this issue of CP violation in the 2HDM, hoping to extract some general insight.

For the purpose of identifying a way to produce charged Higgs bosons, the $W^\mp ZH^\pm$ vertex was studied in $SU(2)\times U(1)$-symmetric gauge models by Grifols and M\'endez in 1980 \cite{Grifols:1980uq}. They pointed out that this coupling resides in the kinetic part of the Lagrangian. Thus, in the Higgs basis, it would be associated with the doublet that has no vacuum expectation value (vev). They showed that if all scalar representations are identical, with hypercharge $Y$ either zero or given by the isospin, $Y=T$, then this coupling vanishes at tree level.
Ten years later, the one-loop contribution to this vertex was evaluated by M\'endez and Pomarol for the MSSM \cite{Mendez:1990epa} and CP-violating effects in the 2HDM were also explored \cite{Mendez:1991gp} (but not for this process).

CP violation in the process
\begin{equation} \label{Eq:int0}
H^\pm(p_1) \to W^\pm(p_2) Z(p_3),
\end{equation}
has recently been explored by Kanemura and Mura \cite{Kanemura:2024ezz} in the 2HDM with explicit CP violation. The asymmetry
\begin{equation} \label{Eq:asymmetry-def}
\delta=\frac{\Gamma(H^+\to W^+ Z)-\Gamma(H^-\to W^- Z)}{\Gamma(H^+\to W^+ Z)+\Gamma(H^-\to W^- Z)},
\end{equation}
arising from one-loop diagrams has been studied.  Their emphasis was on the connection to the custodial symmetry  \cite{Grzadkowski:2010dj}.
While the connection to the custodial symmetry may offer some insight, we think an analysis in terms of physical couplings will add to the understanding of the model, and CP violation in general. We discuss implications of a custodial symmetric 2HDM potential in Section \ref{sect:custodial}.
We shall review this process, and the asymmetry, in terms of physical couplings and masses, allowing for both spontaneous and explicit CP violation.
Actually, while custodial symmetry simplifies some amplitudes and quantitatively impacts the decay rate, it has no qualitative impact: one can have CP conservation without a custodial symmetric potential. Also, one can have a custodial symmetric potential that breaks CP spontaneously, see details in Section \ref{sect:custodial}.

At the one-loop level, the decay (\ref{Eq:int0}) can occur via three distinct mechanisms, three sets of loop diagrams:
(i) Pure bosonic diagrams, no fermions involved;
(ii) Pure fermionic diagrams, fermions couple to the external fields;
(iii) Diagrams where bosons couple to external fields, but the diagrams contain internal fermion (tadpole) loops.

Our focus will be on the decay amplitude ${\cal M}(H^\pm\to W^\pm Z)$, in terms of which the decay rate can be expressed as (all momenta are incoming):
\begin{align}
\Gamma&=\frac{1}{2M}\sum_\text{spin}\int\frac{d^3 p_2}{(2\pi)^32E_2}\frac{d^3 p_3}{(2\pi)^32E_3}|{\cal M}|^2\,(2\pi)^4\delta^{(4)}(p_1+p_2+p_3) \nonumber \\
&=\frac{1}{16\pi M}\sum_\text{spin}\frac{\sqrt{\lambda}}{M^2}\big|{\cal M}\big|^2.
\end{align}
The amplitude ${\cal M}$ carries the dimension of mass and $\lambda$ here refers to  the K\"all\'en function, $\lambda(m_{H^\pm}^2,m_W^2,m_Z^2)$, with $\lambda(x,y,z)=x^2+y^2+z^2-2xy-2yz-2zx$.
Decay rates given in figures in this paper represent an average of the two charge states, $H^+\to W^+ Z$ and $H^-\to W^- Z$.

In order to have CP violation, the overall decay amplitude has to be complex in a non-trivial way (not only due to overall complex coefficients). This means that some loop integrals have to be complex. For this to happen, the invariant mass of an external diagram leg (here: $m_{H^\pm}$) has to exceed the sum of the masses of the two fields of the loop that couple to that particular vertex. We shall see that in the limit of alignment this condition excludes any CP violation in the purely bosonic sector.

It is convenient to split the one-loop amplitude into a bosonic and a fermionic part:
\begin{equation}
{\cal M}={\cal M}^\text{boson} + {\cal M}^\text{fermion},
\end{equation}
where the superscript refers to the loop fields that couple to the external fields $H^\pm$, $W^\pm$ and $Z$. The distinction is not physical, it is just a matter of calculational convenience
due to the involvement of different phase-dependent couplings in the two parts. When the incident $H^\pm$ couples to bosons, the amplitude will be proportional to a coupling referred to as $f_i$ or $f_i^\ast$ (see sections~\ref{sect:Couplings} and \ref{sect:boson}), whereas when it couples to a fermion pair, the amplitude is proportional to the Yukawa coupling $(\tilde\rho^f)^\ast$ (see sections~\ref{sect:Couplings} and \ref{sect:fermions}). 
We include in the ``boson'' amplitude the fermion tadpoles, since these amplitudes are also proportional to a coupling $f_i$ or $f_i^\ast$.

This coupling $f_i$ is a common factor of the trilinear $H^+ H_i W^-$ and the quadrilinear $H^+ H_i ZW^-$ couplings, as well as the corresponding couplings with Goldstone bosons replacing the $W$ or $Z$. The charge-conjugated vertices are proportional to $f_i^\ast$. Since the product $f_i^\ast f_j$ can be related to the more accessible $W^+ W^- H_i$ coupling that we denote $e_i$, this is a convenient way to organise the amplitudes.

Actually, a complex CKM matrix does not contribute to CP violation at the one-loop level of this process, since CKM matrix elements only appear in mutually complex conjugated pairs. But the Yukawa couplings may have other phase-dependent structure, denoted $\tilde\rho$ (see section~\ref{sect:FermionCouplings}), which may contribute to CP violation via the fermionic loop amplitudes.

In general, the different parts will interfere, and will be determined by the structure of the extended scalar sector in which the charged Higgs boson is embedded. 
The asymmetry can then be thought of as consisting of three parts, a bosonic part, a fermionic part, and interference,
\begin{equation}
\delta=\delta^\text{bosonic}+\delta^\text{fermionic}+\delta^\text{interference}.
\end{equation}
 The resulting bosonic CP violation must be related to the CP-odd invariants $\Im J_i$, \cite{Gunion:2005ja,Branco:2005em} ($\Im J_2$ is referred to as $J_1$ in earlier work by Lavoura, Silva and Botella \cite{Lavoura:1994fv,Botella:1994cs}).
 
We shall see that there is a charge asymmetry also in the alignment limit, it originates from the amplitudes of the quark and lepton loops, as well as their interference with the bosonic loops.

Kanemura and Mura \cite{Kanemura:2024ezz} identified CP violation due to interference between the bosonic and fermionic amplitudes. Here, we also find a contribution to the charge asymmetry from CP violation in the purely bosonic sector, as well as from CP violation in the purely fermionic sector. The latter is, however, suppressed by the mass ratios $m_b/m_t$ and $m_\tau/m_t$.

The CPT theorem \cite{Streater:1989vi} dictates that the widths of particle and antiparticle should be the same, $\Gamma_{H^+}=\Gamma_{H^-}$. The asymmetry (\ref{Eq:asymmetry-def}), which originates from the violation of left-right symmetry in the couplings involved, must thus be balanced by an asymmetry in some other channel(s),
\begin{equation} \label{Eq:CPT}
\Gamma(H^+\to W^+Z)+\Gamma(H^+\to \text{other}^+)
=\Gamma(H^-\to W^-Z)+\Gamma(H^-\to \text{other}^-),
\end{equation}
with $\Gamma(H^+\to \text{other}^+) \neq \Gamma(H^-\to \text{other}^-).$
This relation (\ref{Eq:CPT}) must hold, independently of the numerical values of the masses and couplings involved. Thus, we do not expect additional couplings to contribute to $H^\pm \to \text{other}^\pm$.
The matching decays, $H^\pm \to \text{other}^\pm$, can be understood
in terms of the Cutkosky rules  \cite{Cutkosky:1960sp}.
To the one-loop order, these relate the imaginary part of the loop amplitude (responsible for the charge asymmetry) to the decay rate into the final states which would be
\begin{eqnarray}
\text{other}^\pm = \{W^\pm H_i\, (i=1,2,3), W^\pm \gamma, t\bar b/\bar t b\}.\label{eq:decaychannels}
\end{eqnarray}

It is worthwhile noting that in the case of the 2HDM, as well as in any NHDM, there are neither tree level couplings of the form $H^\pm W^\mp Z$ \cite{Grifols:1980uq}, as mentioned before,  nor $H^\pm W^\mp \gamma$ \cite{Gunion:2005ja}. However, the 2HDM allows for the couplings $H_i H^\pm Z_\mu W^{\mp \mu}$ whereas $H_i H^\pm \gamma_\mu W^{\mp \mu}$ are forbidden. This stems from the fact that in a gauge theory such couplings come from the covariant derivative applied to the scalars in the kinetic term. One of the factors in the kinetic term can give rise to $H^\pm W^{\mp \mu}$ and the other to $H_i  Z_\mu$ due to the existence of the term involving the SU(2) generator $T_3$ in the $Z$ coupling.  In the case of the photon its couplings to the scalar doublets are proportional to the electrical charge matrix $Q = T_3 + \frac{1}{2} Y$ which is diagonal with zeros in the lower row. Thus the photon does not couple to neutral scalars. The triple couplings mentioned above would require the neutral gauge boson to couple to a vev. Again this would be impossible for the photon. For the $Z$ boson one cannot obtain such coupling because the kinetic term is composed of the  product of two factors involving the same scalar doublet and the doublet containing  $H^\pm$ has no vev. It is possible to consider extensions of the scalar sector where tree-level couplings of the form $H^\pm W^\mp Z$  could be generated, but this would require the presence of higher representations for the scalars other than doublets  \cite{Grifols:1980uq}.

After setting up our notation for couplings in section~\ref{sect:Couplings} and reviewing the Lorentz structure in section~\ref{sect:Lorentz}, we discuss the contributions to the decay rate and asymmetry from the bosonic and fermionic contributions in sections~\ref{sect:boson} and \ref{sect:fermions}. Interference between these contributions is explored in section~\ref{sect:Boson-fermion}, followed by a discussion of custodial symmetry in section~\ref{sect:custodial} and a summary in section~\ref{sect:summary}.

\section{Couplings and notations}
\label{sect:Couplings}
Rather than expressing the decay rate and asymmetry in terms of potential parameters, vevs and mixing angles (all unphysical), we shall express them in terms of physical couplings. This section presents an overview of the relevant couplings, and also defines our notation for the couplings.
Additional couplings are collected in appendix~\ref{sect:OtherCouplings}.

\subsection{Feynman rules proportional to $e_i$}
\label{sect:GaugeCouplings}

Following Ref.~\cite{Grzadkowski:2014ada}, we list couplings\footnote{All momenta are incoming.} proportional to the factors $e_i$ parametrising the $H_iVV$ couplings in two groups, those\footnote{Right part of eq. (\ref{eq:e_i-epsilon}) corrects a sign error in \cite{Grzadkowski:2014ada}.} that contain the antisymmetric $\epsilon_{ijk}$, and those that do not:
\begin{equation} \label{eq:e_i-epsilon}
H_i H_j Z_\mu: \quad
\frac{g}{2v\cos\thetaW}\epsilon_{ijk}e_k(p_i-p_j)_\mu, \quad
H_i H_j G_0:\quad 
-i\frac{m_i^2-m_j^2}{v^2}\epsilon_{ijk}e_k,
\end{equation}
and
\begin{subequations}
\begin{alignat}{2}
H_i Z_\mu Z_\nu:\quad &
\frac{ig^2}{2\cos^2\thetaW}e_i\,g_{\mu\nu}, &\quad
H_i W^+_\mu W^-_\nu:\quad &
\frac{ig^2}{2}e_i\,g_{\mu\nu}, \label{eq:H_iZZ}\\
H_i G_0 G_0:\quad &
\frac{-im_i^2 e_i}{v^2}, &\quad
H_i G^+ G^-:\quad &
\frac{-im_i^2 e_i}{v^2},\\
H_i G^+ Z_\mu W^-_\nu: \quad &
-\frac{ig^2}{2v} \frac{\sin^2\thetaW}{\cos\thetaW}e_i \,g_{\mu\nu}, &\quad
H_i G^- Z_\mu W^+_\nu: \quad &
-\frac{ig^2}{2v} \frac{\sin^2\theta_\text{W}}{\cos\thetaW}e_i \,g_{\mu\nu}, \\
H_i G_0 Z_\mu: \quad &
\frac{g}{2v\cos\thetaW}e_i(p_i-p_0)_\mu, \\
H_i G^+ W^-_\mu: \quad &
i\frac{g}{2v}e_i(p_i-p^+)_\mu, &\quad
H_i G^- W^+_\mu: \quad &
-i\frac{g}{2v}e_i(p_i-p^-)_\mu.
\end{alignat}
\end{subequations}
Here, $G_0$ and $G^\pm$ denote the would-be Goldstone fields. The factors $e_i$ can be expressed in terms of neutral-sector rotation matrix elements and vevs. For their definition, see \cite{Grzadkowski:2014ada}.
It also follows from the definitions that
\begin{equation} \label{Eq:constraint-e_i}
e_1^2+e_2^2+e_3^2=v^2.
\end{equation}

The trilinear neutral--charged-Higgs coupling is denoted as $q_i$:
\begin{equation}
H_i H^+H^-:\quad 
-iq_i.
\end{equation}

\subsection{Feynman rules proportional to $f_i$ or $f_i^*$}
\label{sect:HpmCouplings}
Some vertices will be proportional to the coupling\footnote{Eq.~(\ref{eq:errorcorrection1}) corrects a misprint in eq. (B.25d) in \cite{Grzadkowski:2014ada}.} $f_i$ parameterising the $H_iH^+ W^-$ coupling (or $f_i^\ast$). These are also expressed in terms of neutral-sector rotation matrix elements and vevs, defined in \cite{Grzadkowski:2014ada} (see also footnote~18 in \cite{Grzadkowski:2018ohf}, where some misprints have been corrected),
\begin{subequations}
\begin{alignat}{2}
H_i H^+ W^{-}_\mu: \quad &
i\frac{g}{2v}f_i(p_i-p_+)_\mu, &\quad
H_i H^-W^{+}_\mu: \quad &
-i\frac{g}{2v}f^*_i(p_i-p_-)_\mu,\label{eq:errorcorrection1}  \\
	H_i H^+ Z_\mu W^{-}_\nu: \quad &
	-\frac{ig^2}{2v} \frac{\sin^2\theta_W}{\cos\theta_W}f_ig_{\mu\nu}, &\quad
	H_i H^- Z_\mu W^{+}_\nu: \quad &
	-\frac{ig^2}{2v} \frac{\sin^2\theta_W}{\cos\theta_W}f^*_ig_{\mu\nu},  \\
	H_iH^+G^-: \quad &
	-\frac{i}{v^2}(m_i^2-m_{H^\pm}^2)f_i, &\quad
	H_iH^-G^+: \quad &
	-\frac{i}{v^2}(m_i^2-m_{H^\pm}^2)f_i^\ast.
\end{alignat}
\end{subequations}

The couplings $f_i$ satisfy the following relations \cite{Grzadkowski:2014ada,Grzadkowski:2018ohf}:
\begin{align}
\Re f_i f_j^\ast&=v^2 \delta_{ij} - e_i e_j, \label{Eq:f_i-real} \\
\Im f_i f_j^\ast&=v \sum_k \epsilon_{ijk} e_k, \\
\sum_{i,j,k} \epsilon_{ijk} e_i f_j^\ast f_k &= -2iv^3, \label{Eq:f_i-cross-product} \\
\sum_i e_i f_i &=0.
\end{align}
Apart from an overall common phase, they are fully determined by the $e_i$ and $v$.

It follows from Eq.~(\ref{Eq:f_i-real}) that the modulus of $f_i$ is given in terms of $e_i$,
\begin{equation} \label{Eq:f_i_modulus}
|f_i|=\sqrt{v^2-e_i^2},
\end{equation}
but, in view of Eq.~(\ref{Eq:f_i-cross-product}), there are some relative complex phases among the $f_i$.
While an overall common phase of the $f_i$ is unphysical, {\it relative} phases are not. 
This equation suggests that, among the bosonic triangle diagrams, those involving the non-SM Higgs field would dominate in the limit of high alignment.
This observation is consistent with the fact that in the Higgs basis, the $H^\pm W^\mp H_i$ coupling originates in the kinetic term for the field $\Phi_2$ that has no vev \cite{Grifols:1980uq,Kanemura:2024ezz}.

If we refer to the degree of alignment for $e_1\to v$ as
\begin{equation} \label{Eq:alignment}
\kappa=\frac{e_1}{v}, \quad\text{with } \kappa\lsim 1,
\end{equation}
then we have
\begin{equation}
|f_1|=v\sqrt{1-\kappa^2}=\sqrt{e_2^2+e_3^2}.
\end{equation}

Away from alignment, one finds
\begin{subequations}
\begin{align}
f_2&=-\frac{f_1(e_1e_2+i e_3 v)}{e_2^2+e_3^2}, \\
f_3&=-\frac{f_1(e_1e_3-i e_2 v)}{e_2^2+e_3^2},
\end{align}
\end{subequations}
whereas in that particular alignment limit ($e_2\to0$ and $e_3\to0$) one has $f_1\to0$, $f_2\to if_3$, ($|f_{2,3}|\to v$), for more details, see eq.~(4.5) of \cite{Grzadkowski:2018ohf}. In the other alignment limits, the couplings are given by cyclic permutations of the above relations. 

\begin{figure}[htb]
\begin{center}
\includegraphics[scale=0.27]{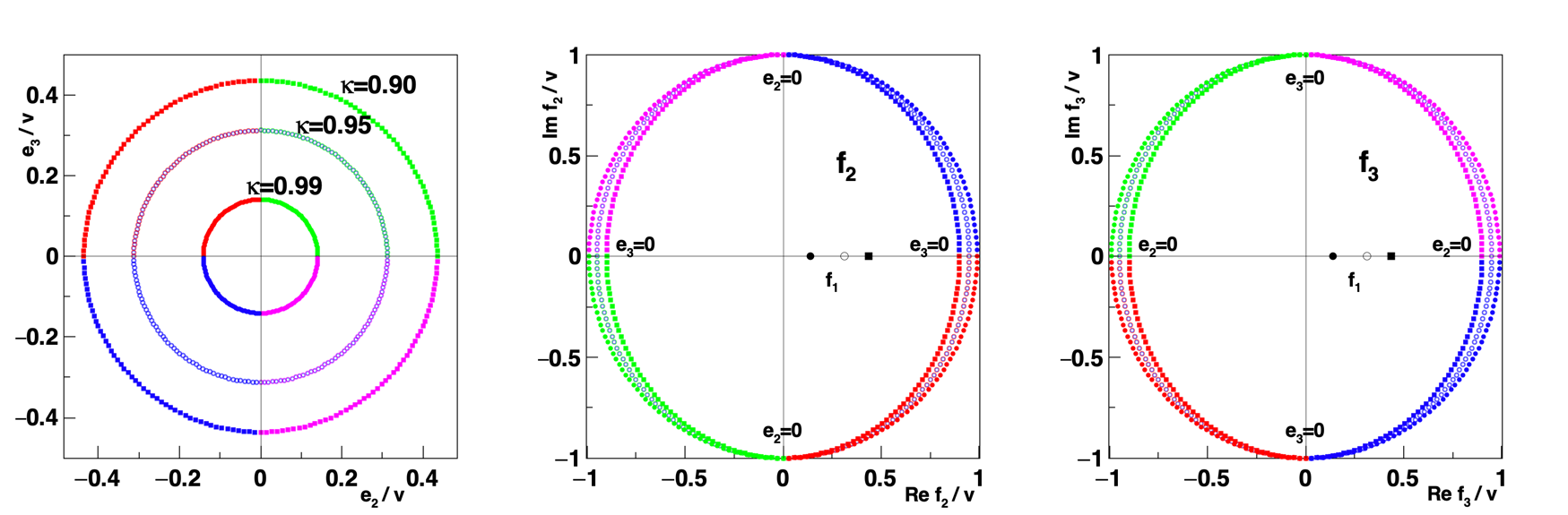}
\end{center}
\vspace*{-4mm}
\caption{Left: circles with constant $e_2^2+e_3^2$; centre and right: charged-Higgs couplings $f_2$ and $f_3$. The phases of the $f_i$ are unphysical, the phases of $f_2$ and $f_3$ are shown with respect to that of $f_1$, which is taken real and positive. Three degrees of alignment, $\kappa$, are considered (see left panel): outer, middle and inner $e_2$-$e_3$ circles refer to $e_1=0.90v$, $e_1=0.95v$, and $e_1=0.99v$, respectively. The corresponding values of $f_1$ are also shown in the middle and right panel. The colours identify the quadrant in the complex $e_2$-$e_3$ plane of the left panel.}
\label{Fig:couplings-e-f}
\end{figure}

For the purpose of gaining some intuition for these couplings $f_i$, we show in Fig.~\ref{Fig:couplings-e-f} how they correlate with $e_2$ and $e_3$ for three cases of high alignment, specified by the $\kappa$ of Eq.~(\ref{Eq:alignment}).
The colour coding refers to the four quadrants in the $e_2$-$e_3$ plane. Notice that going clockwise, the sequence in which the colours appear is the same for $f_2$ and $f_3$ albeit with a shift of 90 degrees.

Interestingly, for a given ratio $e_3/e_2$, the imaginary parts of $f_2$ and $f_3$ are seen to be independent of the degree of alignment, $\kappa$. However, the moduli of the real parts of $f_2$ and $f_3$ are seen to shrink as the alignment is reduced. 

In the bosonic sector, when summing over vector boson polarisations, CP violation will only result in a {\it charge} asymmetry, since the Higgs bosons are Lorentz scalars. 

\subsection{Yukawa couplings involving scalars and fermions}
\label{sect:FermionCouplings}
For the couplings of a neutral scalar to a pair of fermions, we follow the notation of \cite{Haber:2010bw,Grzadkowski:2018ohf}, recalling that $\{u_1,u_2,u_3\}\equiv \{u,c,t\},\ \{d_1,d_2,d_3\}\equiv\{d,s,b\},\ \{l_1,l_2,l_3\}\equiv\{e^-,\mu^-,\tau^-\}$. Diagonal couplings are given by 
\begin{subequations} \label{Eq:Scalar-fermion}
\begin{alignat}{2}
&H_j \bar u_k u_k: &\quad &\frac{-im_{u_k}}{v^2}\, e_j
-\frac{i}{\sqrt2\,v}\left[(\tilde\rho_{kk}^U)^\ast P_L f_j^\ast+ \tilde\rho_{kk}^U P_R f_j \right], \\
&H_j \bar d_k d_k: &\quad &\frac{-im_{d_k}}{v^2}\, e_j
-\frac{i}{\sqrt2\,v}\left[(\tilde\rho_{kk}^D)^\ast P_R f_j^\ast+ \tilde\rho_{kk}^D P_L f_j \right], \\
&H_j \bar \ell_k \ell_k: &\quad &\frac{-im_{\ell_k}}{v^2}\, e_j
-\frac{i}{\sqrt2\,v}\left[(\tilde\rho_{kk}^L)^\ast P_R f_j^\ast+ \tilde\rho_{kk}^L P_L f_j \right],
\end{alignat}
\end{subequations}
with $P_{L/R}=\half(1\mp\gamma_5)$ the projection operators. For leptons, non-diagonal couplings would imply violation of lepton number, not experimentally observed. We do not present non-diagonal couplings here, but refer the reader to Section E.2 of \cite{Grzadkowski:2018ohf}.
The superscripts $U$, $D$ and $L$ refer to up- and down-type quarks and charged leptons, respectively.
Note that whereas we have extracted the physical couplings $f_j$ and $f_j^\ast$, some authors \cite{Aoki:2009ha,Branco:2011iw} extract a factor $m/v$ from the definition of couplings analogous to $\tilde\rho$.
In view of the strong mass hierarchy among fermions, we shall also refer to the dominant terms $\tilde\rho^U_{33}$, $\tilde\rho^D_{33}$ and $\tilde\rho^L_{33}$ as $\tilde\rho_t$, $\tilde\rho_b$ and $\tilde\rho_\tau$.

\def\qq{\hspace*{3mm}}
For numerical illustrations, we only consider diagonal $\tilde{\rho}$ matrices. 
The charged-Higgs and Goldstone-Yukawa couplings then become
\begin{subequations}\label{Eq:Yukawas}
\begin{alignat}{4}
&H^+ d_k\bar u_m: &\qq &iV_{mk}[(\tilde\rho^U_{mm})^\ast P_L-(\tilde\rho^D_{kk})^\ast P_R], &\quad
&H^- \bar d_k u_m: &\qq &iV_{mk}^\ast[\tilde\rho^U_{mm} P_R-\tilde\rho^D_{kk} P_L], \\
&H^+ \ell_k\bar \nu_k: &\qq &-i(\tilde\rho^L_{kk})^\ast P_R, &\quad
&H^- \bar\ell_k \nu_k: &\qq &-i\tilde\rho^L_{kk} P_L, \\
&G^+ d_k\bar u_m: &\qq & \frac{i\sqrt2\, V_{mk}}{v} [m_{u_m} P_L-m_{d_k} P_R], &\quad
&G^- \bar d_k u_m: &\qq & \frac{i\sqrt2\, V_{mk}^\ast}{v} [m_{u_m} P_R-m_{d_k} P_L], \\
&G^+ \ell_k\bar \nu_k: &\qq & \frac{-i\sqrt2}{v} m_{l_k} P_R, &\quad
&G^- \bar\ell_k \nu_k: &\qq & \frac{-i\sqrt2}{v} m_{l_k} P_L.
\end{alignat}
\end{subequations}
Here, $V$ is the CKM matrix.
We recall that there will be a factor $V_{mk}$ or $V_{mk}^\ast$ associated with the coupling of the fermions to the $W^\pm$ and note that the $V_{mk}$ factors come in conjugated pairs, cancelling any dependence on a complex phase. The dominant couplings involving top/bottom quarks and tau leptons are obtained by putting $k=m=3$.

The Yukawa couplings are in the Type~I and Type~II models taken proportional to the fermion mass, but $\tilde\rho$ is in general independent of the fermion masses.

\subsection{Couplings involving gauge bosons and fermions}
\label{sect:YukawaCouplings}
We expand the couplings of the gauge bosons $W^\pm$ and $Z$ in terms of Left- and Right-handed parts,
\begin{alignat} {2}
&W^+_\mu l \bar \nu_l: \quad &&-\frac{ig}{\sqrt2}\gamma^\mu P_L \equiv a_W \gamma^\mu P_L, \\
&W^-_\mu \bar l \nu_l: \quad && -a_W \gamma^\mu P_L, \\
&W^+_\mu \bar u_m d_k: \quad && a_W V_{mk}\gamma^\mu P_L, \\
&W^-_\mu u_m \bar d_k: \quad && -a_W V_{mk}^*\gamma^\mu P_L, \\
&Z_\mu \bar f f: \quad && \frac{-ig}{2\cos\theta_W}\gamma^\mu[Z^f_L P_L + Z^f_R P_R]\equiv a_Z\gamma^\mu[Z^f_L P_L + Z^f_R P_R],
\end{alignat}
with the product
\begin{equation}
a_W a_Z=-\frac{g^2}{2\sqrt2\cos\theta_W}.
\end{equation}
Then, we have for the quarks
\begin{subequations} \label{Eq:Z-couplings}
\begin{alignat}{2}
&Z^u_L= 1-(4/3)\sin^2\theta_W, &\quad &Z^u_R= -(4/3)\sin^2\theta_W, \\
&Z^d_L= -1+(2/3)\sin^2\theta_W, &\quad &Z^d_R= (2/3)\sin^2\theta_W, 
\end{alignat}
\end{subequations}
and for the leptons
\begin{subequations}
\begin{alignat}{2}
&Z^\nu_L= 1, &\quad &Z^\nu_R= 0, \\
&Z^\ell_L= -1+2\sin^2\theta_W, &\quad &Z^\ell_R= 2\sin^2\theta_W.
\end{alignat}
\end{subequations}

\subsection{Phases}
\label{sect:Phases}

Since CP violation can be formulated in terms of the interplay of complex quantities, it is useful to present an overview of the different complex quantities relevant to the considered process:
\begin{itemize}
\item
Loop integrals become complex (we exclude from this discussion the overall $i$ from rotation to, and integration in Euclidean space) when the invariant mass of the external field at a particular vertex exceeds the sum of the masses of the two particles in the loop lines at that vertex. In terms of LoopTools notation \cite{Hahn:1998yk}, the loop integral becomes complex when $\sqrt{s_i}>m_i+m_{i+1}$ for some $i$. Thus, if a diagram has an internal $H^\pm$ field connected to the incoming $H^\pm$ field, the loop integral remains real;
\item
The bosonic amplitudes are proportional to the complex couplings $f_i$ (or $f_i^\ast$). Relative phases between the $f_i$ (as illustrated in Fig.~\ref{Fig:couplings-e-f}) are physical, given in terms of the couplings $e_j$ and play a role, whereas the overall phase of the $f_i$ plays no role for these amplitudes;
\item
The fermionic amplitudes are proportional to the complex quantities $\tilde\rho$. Again, phase differences between the couplings of different fermions, $\tilde\rho^U$, $\tilde\rho^D$ and $\tilde\rho^L$ are physical and play a role for the overall fermionic amplitudes. The absolute phase plays no role for these amplitudes;
\item
The CKM matrix elements involved in vertices with a charged boson ($W^\pm$, $H^\pm$ or $G^\pm$) and a fermion--antifermion pair, are in general complex. However, any $u_m\bar d_k$ pair created at one vertex will have to annihilate at another, hence these CKM phases cancel.
\item
Finally, the bosonic and fermionic amplitudes may interfere. This interference depends on the ``overall'' reference phases in the bosonic ($f_i$) and fermionic ($\tilde\rho$) sectors, and is most easily illustrated by the product $f_i\tilde\rho^f$ appearing in the fermionic tadpole amplitudes of table~\ref{Table:fermion-tadpole-amplitudes}. 

With three complex couplings $f_i$ in the bosonic sector (associated with the three neutral states $h_i$) and three complex couplings $\tilde\rho^f$ in the fermion sector (associated with up- and down-type quarks and charged leptons), we have a total of six phases. Only differences are physical, so the number of physical phases (arising from these fields) is five. 
Rather than picking five relative phases between fermion and boson fields, we define two relative phases in the bosonic sector (phases of $f_2$ and $f_3$ relative to the reference phase of $f_1$), two in the fermionic sector (phases of $\tilde\rho^U$ and $\tilde\rho^D$ with respect to $\tilde\rho^L$) and one phase relating $\tilde\rho^L$ to $f_1$.

In our terminology, the latter physical phase will appear as a sum of the phases of $f_1$ and $\tilde\rho^L$ (see the example $f_i\tilde\rho^f$ mentioned above), not a difference. Concretely, we take $f_1$ real, and shift the reference phase in the fermionic sector, namely that of $\tilde\rho_{33}^L$ by $\theta_\text{BF}$, $\zeta_\tau\equiv\theta_\text{BF}$, where $\zeta_\tau$ is the phase of $\tilde\rho_{33}^L$. The phases of $\tilde\rho_{33}^D$ and $\tilde\rho_{33}^U$, namely $\zeta_b$ and $\zeta_t$, will be given with respect to $\zeta_\tau$.
\end{itemize}

\subsection{Parameter restrictions}
\label{sect:Restrictions}

\paragraph{Close to Alignment.}
The discovered 125~GeV Higgs boson has a coupling to $W^+W^-$ and to $ZZ$ very close to that expected for the Standard Model. The SM-limit is referred to as alignment. We shall assume that the discovered state is the lightest one, $h_1$. In the adopted terminology, near-alignment then means $e_1\simeq v$ (see section~\ref{sect:HpmCouplings}). From Eq.~(\ref{Eq:constraint-e_i}) it follows that
\begin{equation}
e_2^2+e_3^2\ll v^2,\quad\text{or}\quad \sqrt{e_2^2+e_3^2}\ll e_1.
\end{equation}

\paragraph{Close to CP conservation.}
The discovered 125~GeV Higgs boson has couplings compatible with it  being a CP-even state. Importantly, that does not imply that (contrary to what is often assumed) among the three neutral scalars there are two even ones and one that is odd under CP.
Because of near-alignment, $h_2$ and $h_3$ will only couple weakly to $W^+W^-$ and to $ZZ$.
As mentioned above, it does not follow that $h_2$ or $h_3$ should have a small coupling to $h^+h^-$, given by $q_2$ and $q_3$. In terms of the scalar-sector invariants given in appendix~\ref{sect:CP-2HDM}, the quantity $\Im J_{30}$ is not constrained.

In the SM, CP violation only comes from the fermionic sector and requires a complex $V_\text{CKM}$. It is by now established that $V_\text{CKM}$ must be complex \cite{Botella:2005fc,Charles:2004jd}.
In the 2HDM, other sources of CP violation may appear in the interaction between fermions and bosons and would require relative $\zeta$ phases different from zero.

\paragraph{Boundedness from below and unitarity.}
A simple sufficient (but not {\it necessary}) condition for boundedness from below of the vacuum can be expressed as \cite{OMO24}:
\begin{align}
D&=\frac{-1}{4v^{10}}\left[ m_2^2 m_3^2(v^2q_1-2e_1m_{H^\pm}^2)^2
+m_3^2 m_1^2(v^2q_2-2e_2m_{H^\pm}^2)^2
+m_1^2 m_2^2(v^2q_3-2e_3m_{H^\pm}^2)^2 \right] \nonumber \\
&+\frac{q\, m_1^2 m_2^2 m_3^2 }{2v^6}>0,
\end{align}
where $q$ is the quartic $H^+H^+ H^- H^-$ coupling, whereas the others have been defined above.
For typical parameters, with $m_1^2<m_2^2<m_3^2$ and $e_1={\cal O}(v)$ (in order to be near alignment), the first term in the square bracket tends to dominate unless $q_1$ is also taken ``large'' such that $|v^2q_1-2e_1m_{H^\pm}^2|$ is small.

The coupling $q$ does not enter in our calculation, so one might be tempted to satisfy boundedness from below by taking $q$ large. However, this might violate the unitarity constraint, which requires $q<4\pi$.
At low masses, $m_{H^\pm}$, not higher than around $350~\text{GeV}$, the above criterion is satisfied. At higher masses, one needs to explore further constraints involving the eigenvalues of the matrix $\Lambda_E$ of the bilinear formalism \cite{Ivanov:2015nea}. One step in this procedure involves checking whether any of its eigenvalues is complex. This can be achieved by expressing $\Lambda_E$ in terms of physical couplings \cite{OMO24}.

\section{Lorentz structure}
\label{sect:Lorentz}
In general, the amplitude ${\cal M}(H^\pm\to W^\pm Z)^{\alpha\beta}$ has a rich Lorentz structure. However, we shall assume the $W^\pm$ and $Z$ will be observed via their decays to light fermions, this allows us to drop terms proportional to either $p_2^\alpha$ or $p_3^\beta$. Thus, with $r$ and $s$ denoting the final-state polarizations,
\begin{equation}
{\cal M}_{rs}={\cal M}_{rs}^{\alpha\beta} \epsilon_{r\alpha} \epsilon_{s\beta},
\end{equation}
we have\footnote{We will often use a ``light'' notation, suppressing indices $r$ and $s$.} \cite{Mendez:1990epa}
\begin{equation} \label{Eq:expansion}
{\cal M}^{\alpha\beta}=F\,g^{\alpha\beta} + \frac{G}{m_W^2} p_3^\alpha p_2^\beta
+\frac{H}{m_W^2} \epsilon^{\alpha\beta}_{\ \ \mu\nu}p_3^\mu p_2^\nu, 
\end{equation}
where the $H$ term represents parity violation. 

There are some constraints on which diagrams will contribute to which amplitude:
\begin{itemize}
\item
$G$ only gets contributions from triangle diagrams (having two separate ``connections'' to the final-state vector bosons), no ``bubbles'' or tadpoles;
\item
$H$ only gets contributions from triangle diagrams involving fermions.
\end{itemize}

With the polarization sum ($m_V\in\{m_W,m_Z\}$)
\begin{equation}
\sum_r \epsilon_r^\mu(p) \epsilon_r^\nu(p) = - g^{\mu\nu}+\frac{p^\mu p^\nu}{m_V^2},
\end{equation}
and the abbreviations
\begin{align}
\xi&=\frac{p_2\cdot p_3}{m_W^2}=\frac{s_1-m_W^2-m_Z^2}{2m_W^2}, \\
(s_1-m_W^2-m_Z^2)^2&=\lambda(s_1,m_W^2,m_Z^2)+4m_W^2 m_Z^2,
\end{align}
where (with all particles on-shell)
\begin{equation} \label{Eq:s_i}
s_1=p_1^2=m_{H^\pm}^2, \quad s_2=p_2^2=m_{W^\pm}^2, \quad s_3=p_3^2=m_Z^2,
\end{equation}
we can write out the polarization sums in compact form, for the interference of diagrams (d) and (d'):
\begin{align} \label{Eq:spin-sum}
\sum_{r,s}{\cal M}_{rs}^\text{(d)} {\cal M}_{rs}^{\text{(d')}\ast}
&=\left(3+\frac{\lambda}{4m_W^2 m_Z^2}\right)F^\text{(d)} F^{\text{(d')}\ast}
+\frac{\lambda\xi}{4m_W^2 m_Z^2}
\left[F^\text{(d)} G^{\text{(d')}\ast}+G^\text{(d)} F^{\text{(d')}\ast}\right]
\nonumber \\
&+\frac{\lambda^2}{16m_W^6 m_Z^2} G^\text{(d)} G^{\text{(d')}\ast}
+\frac{\lambda}{2m_W^4} H^\text{(d)} H^{\text{(d')}\ast}.
\end{align}
We note that the $H$ amplitude, which originates from fermionic triangle diagrams, does not interfere either with the amplitude $F$ or with $G$.

\section{Bosonic contributions}
\label{sect:boson}

The bosonic diagrams for $H^+\to W^+Z$ all contain a factor $f_i$. In fact, it is convenient to include diagrams with fermion tadpoles in this category, since they also involve the coupling $f_i$, in contrast to diagrams where the incoming $H^\pm$ couples directly to a fermion--antifermion pair.

Let us parametrize the $F$- and $G$-amplitudes of the bosonic loop diagrams as
\begin{subequations} \label{Eq:boson:H+}
\begin{eqnarray}
F^\dd&=&N_0\,K_\dd\, f_i\, {\cal F}_i^\dd \label{famplitude},\\
G^\dd&=&N_0\, K_\dd\,  f_i\, {\cal G}_i^\dd.
\end{eqnarray}
\end{subequations}
Here, the index `d' refers to a set of Feynman diagrams, the index $i$ which should be summed over from 1 to 3 labels the neutral scalars, and\footnote{In $N_0$, ``$i$'' is the imaginary unit.}
\begin{equation}
N_0=\frac{i}{16\pi^2}.
\end{equation}

The amplitudes of the corresponding charge-conjugated processes are given by
\begin{subequations} \label{Eq:boson:H-}
\begin{eqnarray}
\tilde{F}^\dd&=&N_0\,K_\dd^*\, f_i^*\, {\cal F}_i^\dd ,\label{conjugatefamplitude}\\
\tilde{G}^\dd&=&N_0\, K_\dd^*\, f_i^*\, {\cal G}_i^\dd.
\end{eqnarray}
\end{subequations}
As mentioned above, there is no $H$-amplitude for bosonic loop diagrams.

The complex functions ${\cal F}_i^\dd$ and ${\cal G}_i^\dd$ refer to the loop integral for the diagrams `d'.
These integrals are the {\it same} for both charge configurations, whereas the prefactor $K_{d}\, f_i$ undergoes complex conjugation as will be discussed in section~\ref{sect:bosondiagrams}.

CP violation arises when the two moduli on the left-hand side of Eqs.~(\ref{Eq:boson:H+}) and (\ref{Eq:boson:H-}) are different. This will happen when the complex phases arising from different diagrams are unequal in the two sums. In contrast, if the terms in the two sums were pairwise related by complex conjugation, there would be no CP violation.

\subsection{Fermion tadpoles}
\label{sect:fermions-tadpoles}

We shall discuss fermion tadpoles separately, since they involve a non-trivial (and phase-dependent) coupling between a neutral scalar and a fermion-antifermion pair.
Because of the overall factors of $f_i$ and $f_i^\ast$, we decompose these like the bosonic amplitudes of Eqs.~(\ref{famplitude}) and (\ref{conjugatefamplitude}).

For the process considered, these couplings will only be relevant for the one-point functions, and thus the $\gamma_5$ part plays no role. 

\subsection{Decay via bosonic loops}
The bosonic loop diagrams only contribute to the $F$ and $G$ amplitudes. For a schematic discussion of the resulting CP violation, let us define in this case
\begin{equation} \label{Eq:L-define}
{\cal M}^{\alpha\beta}=\sum_i \sum_\dd f_i\, A_i^{\text{(d)}\alpha\beta},
\end{equation}
with
\begin{equation} \label{Eq:L-define-2}
A_i^{\text{(d)}\alpha\beta}=F_i^\text{(d)}\,g^{\alpha\beta}+\frac{G_i^\text{(d)}}{m_W^2}\,p_3^\alpha\, p_2^\beta 
=N_0K_\dd\left[{\cal F}_i^\dd \,g^{\alpha\beta}+{\cal G}_i^\dd\frac{p_3^\alpha\, p_2^\beta}{m_W^2}\right].
\end{equation}

The $H^\pm W^\mp H_i$ coupling $f_i$ is in general complex, and this phase is independent of the complex phases of the loop integrals. Schematically, we can re-write the one-loop bosonic amplitudes as
\begin{subequations}
\begin{align}
{\cal M}(H^+\to W^+Z)^{\alpha\beta}&=\sum_{i}\sum_\dd f_i A_i^{\text{(d)}\alpha\beta}, \\
{\cal M}(H^-\to W^-Z)^{\alpha\beta}&=\sum_{i}\sum_\dd f_i^\ast A_i^{\text{(d)}\alpha\beta},
\end{align}
\end{subequations}
where we have suppressed the polarisation indices $r$ and $s$, and where the dependences on $f_i$ and $f_i^\ast$ are exhibited.

In order to see how the charge asymmetry arises, let us consider the first term, $F^{(\dd)} F^{(\dd')}{}^\ast$, in Eq.~(\ref{Eq:spin-sum}), summed over the diagrams,
\begin{subequations} \label{Eq:a-double-sum}
\begin{alignat}{2} 
&H^+\to W^+Z: &\quad \sum_{d,d'} F^{(\dd)} F^{(\dd')}{}^\ast &=|N_0|^2 \sum_{i,i'} f_i f_{i'}^\ast \sum_{d,d'}
(K_\dd {\cal F}_i^\dd) (K_{\dd'}^\ast {\cal F}_{i'}^{\dd'}{}^\ast), \label{Eq:H+decay}\\
&H^-\to W^-Z: &\quad \sum_{d,d'} \tilde F^{(\dd)} \tilde F^{(\dd')}{}^\ast&=|N_0|^2 \sum_{i,i'} f_i^\ast f_{i'} \sum_{d,d'}
(K_{\dd}^\ast {\cal F}_i^\dd )(K_{\dd'}{\cal F}_{i'}^{\dd'}{}^\ast). \label{Eq:H-decay}
\end{alignat}
\end{subequations}

Consider for a moment the case of all $K_\dd$ being real (they are not). Then the two expressions would only differ by the factor $(f_i f_{i'}^\ast)$ in Eq.~(\ref{Eq:H+decay}) being complex conjugated in Eq.~(\ref{Eq:H-decay}). In this case, for a given $i$, the moduli of $\sum_\dd K_\dd {\cal F}_i^\dd$ and $\sum_\dd K^\ast_\dd {\cal F}_i^\dd$ would be the same.

However, since the factor $(f_i f_{i'}^\ast)$ is complex (as discussed in section~\ref{sect:HpmCouplings}), and since some loop integrals ${\cal F}_i^\dd$ will in general be complex, the adding over $i$ (or $i'$) would give a different result for the two cases of Eq.~(\ref{Eq:a-double-sum}). Thus, the fact that not all loop amplitudes ${\cal F}_i^\dd$ (or all ${\cal G}_i^\dd$) have the same complex phase will contribute to a charge asymmetry (there will also be independent contributions from the fermionic loops).

For the pure bosonic amplitudes, the dependence on the couplings $f_i$ and $f_{i^\prime}^\ast$ can be represented in terms of a dependence on the more familiar $W^+W^-H_i$ coupling $e_i$, as discussed in section~\ref{sect:HpmCouplings}. 

\subsection{Bosonic diagrams}
\label{sect:bosondiagrams}
We split the discussion of bosonic amplitudes into three parts: (1) triangle diagrams, (2) bubble and tadpole diagrams, and (3) fermionic tadpole diagrams. These all have an overall factor of $f_i$ or $f_i^\ast$.
\subsubsection{Triangle diagrams}

\begin{figure}[htb]
\begin{center}
\includegraphics[scale=0.5]{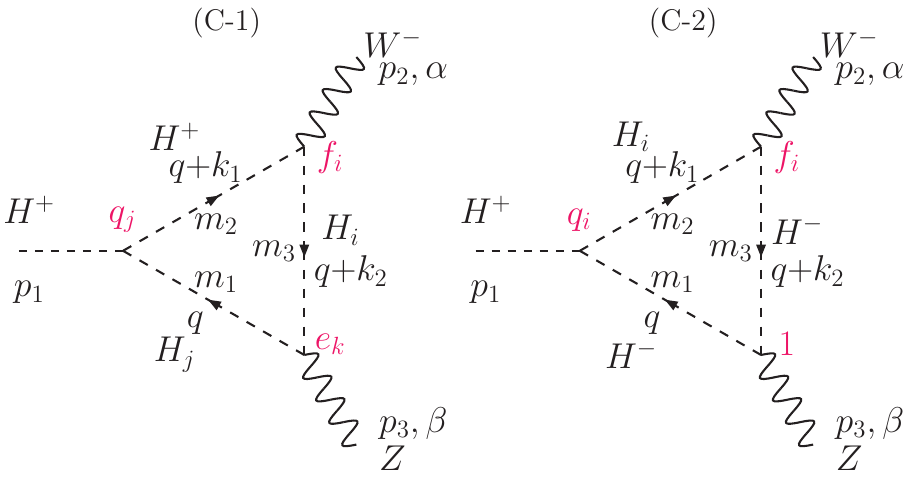}
\end{center}
\vspace*{-4mm}
\caption{Boson triangle diagrams involving the $q_i$ (or $q_j$) couplings.}
\label{Fig:feynman-a}
\end{figure}

\begin{figure}[htb]
\begin{center}
\includegraphics[scale=0.5]{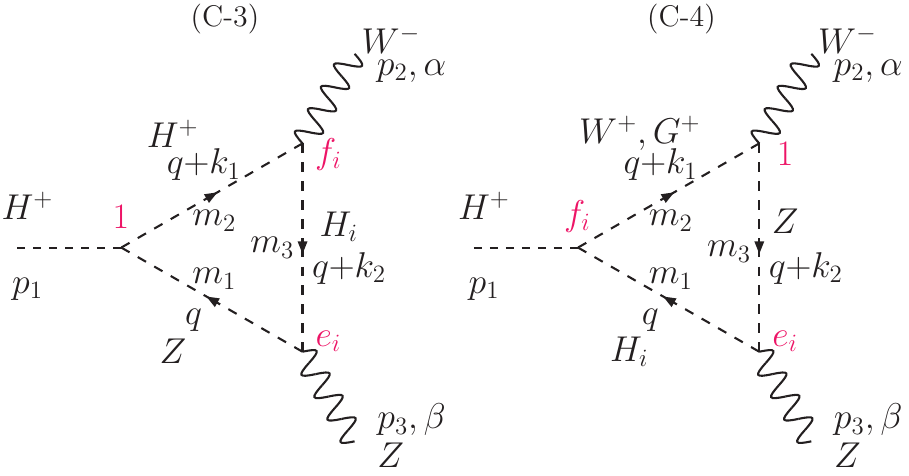} \
\includegraphics[scale=0.5]{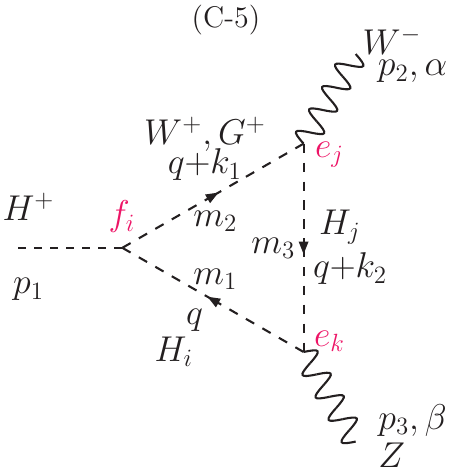} \\
\includegraphics[scale=0.5]{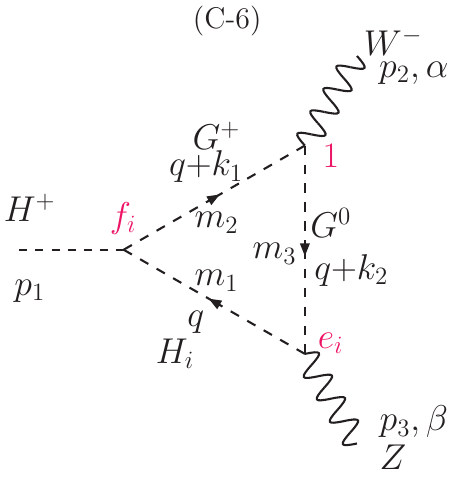} \
\includegraphics[scale=0.5]{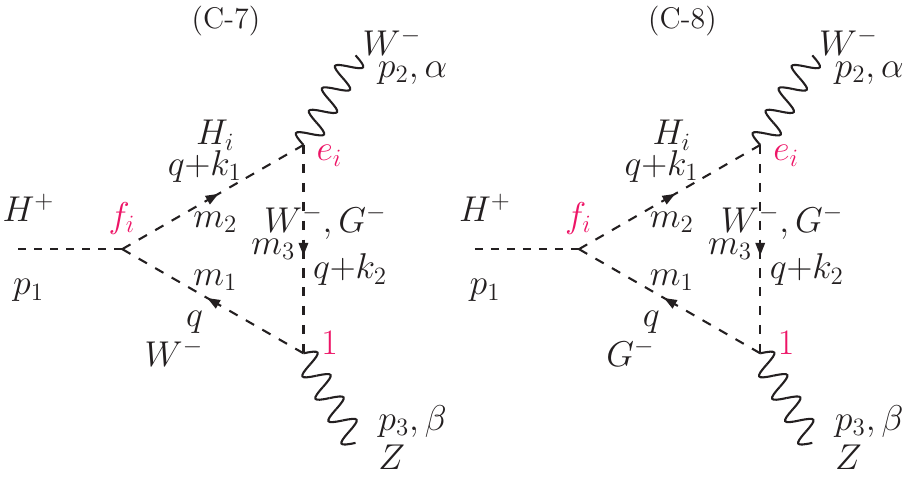}
\end{center}
\vspace*{-4mm}
\caption{Other boson triangle diagrams. Some represent two cases, with a $W^\pm$ or a $G^\pm$.}
\label{Fig:feynman-b}
\end{figure}

In Figs.~\ref{Fig:feynman-a} and \ref{Fig:feynman-b} we show triangle diagrams involving bosonic fields.
The notations for internal momenta and masses follow the LoopTools convention \cite{Hahn:1998yk}, whereas the non-trivial parts of the couplings (in red) follow the convention of Ref.~\cite{Grzadkowski:2014ada}.

The couplings involved in these diagrams are of two kinds. Trilinear couplings among scalars originate from the potential, and are of ${\cal O}(1)$. Couplings involving gauge fields $W$ or $Z$ come from the kinetic terms, and carry one gauge coupling $g$ for each gauge boson. However, terms of relative order $g^2$ are not necessarily of relative order $\alpha$, since an accompanying denominator $4\pi$ may be lacking.

In order to exhibit our notation, which is given by Eqs.~({\ref{Eq:boson:H+}) and ({\ref{Eq:boson:H-}), we write out explicitly the amplitudes for the Feynman diagrams involving $q_i$ (or $q_j$) as
\begin{subequations} \label{Eq:diagrams-a}
\begin{align}
A_i^{\text{(C-1)}\alpha\beta}&=\frac{ig^2}{4v^2\cos\theta_W}\sum_{j,k}\epsilon_{ijk} q_j e_k
\int \frac{d^4q}{(2\pi)^4} \frac{(2q+k_1+k_2)^\alpha (2q+k_2)^\beta}
{[q^2-m_j^2][(q+k_1)^2-m_{H^\pm}^2][(q+k_2)^2-m_i^2]}, \\
A_i^{\text{(C-2)}\alpha\beta}&=\frac{g^2\cos2\theta_W}{4v\cos\theta_W}q_i
\int \frac{d^4q}{(2\pi)^4} \frac{(2q+k_1+k_2)^\alpha (2q+k_2)^\beta}
{[q^2-m_{H^\pm}^2][(q+k_1)^2-m_i^2][(q+k_2)^2-m_{H^\pm}^2]},
\end{align}
\end{subequations}
where $A_i^{\text{(d)}\alpha\beta}$ is defined by Eqs.~(\ref{Eq:L-define}) and (\ref{Eq:L-define-2}).
Some of the remaining triangle diagrams cover two cases, both with a $W^\pm$ field and with a $G^\pm$ field in the loop.

\begin{table}[htb]
	\caption{Amplitudes for triangular bosonic diagrams}
	\label{Table:boson-C-amplitudes}
	\vspace*{2mm}
	\resizebox{\columnwidth}{!}{
		\begin{tabular}[b]{|c|c|p{6.7cm}|c|}
			\hline
			d & $K_\dd$ &  \hspace*{3cm}${\cal F}_i^{\dd}$ &  ${\cal G}_i^{\dd}$  \\
			\hline
			C-1& $\frac{i  g^2 }{v^2 c_{\text{W}}}$  & $\displaystyle\sum_{j,k}\epsilon_{ijk}q_je_k C_{00}$ & $\displaystyle\sum_{j,k}\epsilon_{ijk}q_j e_k  \left(C_{\text{1}}+C_{\text{11}}+C_{\text{12}}\right)$  \\
			\hline
			C-2& $\frac{g^2 c_{\text{2W}}}{v c_{\text{W}}}$  & $q_i C_{00}$ & $q_i\left(C_{\text{1}}+C_{\text{11}}+C_{\text{12}}\right)$  \\
			\hline
			C-3& $\frac{ g^4 c_{\text{2W}}}{4 v c_{\text{W}}^3}$  & $ e_i C_{00}$ & $e_i\left(2 C_0+3 C_1+2 C_2+C_{11}+C_{12}\right)$  \\
			\hline
			\multirow{4}*{C-4W}& \multirow{4}*{$\frac{ g^4}{4 v c_{\text{W}}}$}  & $e_i (\frac{1}{2}-3 C_{00}
			$ \par 
			$\hspace*{0.5cm}+ s_1 \left(C_1+C_2-C_{11}-C_{12}\right)$ \par 
			$\hspace*{0.5cm}- s_2 \left(C_0-C_1+C_2-C_{12}\right)$ \par 
			$\hspace*{0.5cm}+ s_3 \left(C_0-C_1-C_2-C_{12}-C_{22}\right))$
			& \multirow{4}*{$e_i\left(-C_1-4 C_2+C_{11}+C_{12}\right)$}  \\
			
			\hline
			C-4G& $\frac{ g^4  s_{\text{W}}^2}{4 v c_{\text{W}}^3}$  & $e_i \left(m_i^2-m_{{H^\pm}}^2\right) C_0$ & $0$  \\
			\hline
			C-5W& $\frac{i  g^4}{4 v^2 c_{\text{W}}}$  & $\displaystyle\sum_{j,k}\epsilon_{ijk}e_j e_kC_{00}$ & $\displaystyle\sum_{j,k}\epsilon_{ijk}e_j e_k(-C_1+C_{11}+C_{12})$  \\
			\hline
			C-5G& $-\frac{i  g^2 }{v^4 c_{\text{W}}}$  & $\displaystyle\sum_{j,k}\epsilon_{ijk}e_j e_k\left(m_i^2-m_{{H^\pm}}^2\right)C_{00}$ & $\displaystyle\sum_{j,k}\epsilon_{ijk}e_j e_k\left(m_i^2-m_{{H^\pm}}^2\right)(C_1+C_{11}+C_{12})$  \\
			\hline
			C-6& $-\frac{ g^2 }{v^3 c_{\text{W}}}$  & $e_i\left(m_i^2-m_{{H^\pm}}^2\right)C_{00}$ & $e_i\left(m_i^2-m_{{H^\pm}}^2\right)(C_1+C_{11}+C_{12})$  \\
			\hline
			\multirow{4}*{C-7W}& \multirow{4}*{$-\frac{ g^4 c_{\text{W}}}{4 v}$}  & $e_i(\frac{1}{2}-3 C_{00}$\par 
			$\hspace*{0.5cm}- s_1 \left(2 C_0+3 C_1+C_2+C_{11}+C_{12}\right)$\par 
			$\hspace*{0.5cm}+ s_2 \left(2 C_0+C_1+C_2+C_{12}\right)$\par 
			$\hspace*{0.5cm}- s_3 \left(2 C_0+C_1+3 C_2+C_{12}+C_{22}\right))$
			& \multirow{4}*{$e_i \left(2 C_0+3 C_1-2 C_2+C_{11}+C_{12}\right)$}  \\
			\hline
			C-7G& $\frac{ g^4 s_{\text{W}}^2}{4 v c_{\text{W}}}$  & $e_iC_{00}$ & $e_i(2 C_0+3 C_1+2 C_2+C_{11}+C_{12})$  \\
			\hline
			C-8W& $\frac{ g^4  s_{\text{W}}^2}{4 v c_{\text{W}}}$  & $e_i\left(m_i^2-m_{{H^\pm}}^2\right)C_0$ & $0$  \\
			\hline
			C-8G& $\frac{ g^2 c_{\text{2W}} }{v^3 c_{\text{W}}}$  & $e_i\left(m_i^2-m_{{H^\pm}}^2\right)C_{00}$ & $e_i\left(m_i^2-m_{{H^\pm}}^2\right)(C_1+C_{11}+C_{12})$  \\
			\hline
		\end{tabular}}
\end{table}

The results for the loop integrals ${\cal F}_i^\dd$ and ${\cal G}_i^\dd$ are given in table~\ref{Table:boson-C-amplitudes} together with the prefactor $K_\dd$ in the notation of Eq.~(\ref{Eq:boson:H+}). The $C$'s are the triangle loop integrals, using the notation of Ref.~\cite{Hahn:1998yk} and can be expressed in terms of Passarino--Veltman functions \cite{Passarino:1978jh}.   We recall that $i\in\{1, 2, 3\}$ labels the neutral scalar\footnote{Exception: in the $K_\dd$ column, `$i$' denotes the imaginary unit.} that couples to $H^+$ (or $H^-$), whereas `d' labels the diagram.

Numerical codes are available \cite{vanOldenborgh:1989wn,Hahn:1998yk}. The following facts are important to bear in mind: (1) Since the coupling vanishes at tree level, the overall loop amplitude is finite; (2) Some of these integrals are divergent, so infinities have to be subtracted consistently. For this purpose, a tool like FormCalc \cite{Hahn:2000kx,Kublbeck:1990xc,Hahn:1998yk} is very valuable.

In table~\ref{Table:boson-C-amplitudes}, we do not specify the arguments of the triangle loop integrals, they will always be $C_X\equiv C_X(s_1,s_2,s_3,m_1^2,m_2^2,m_3^2)$, where the $s_i$ are given by Eq.~(\ref{Eq:s_i}),
whereas $m_1$, $m_2$ and $m_3$ refer to the masses of the propagators involved, starting with the lower one (between the external $Z$ and $H^\pm$), and proceeding clockwise. They must be identified for each diagram by comparing with the corresponding Feynman diagram, and should not be confused with the masses of the three neutral states, referred to by the same symbols, $m_1$, $m_2$ and $m_3$.

Further remarks:
\begin{itemize}
\item
These results have been obtained by FormCalc  \cite{Hahn:2000kx,Kublbeck:1990xc,Hahn:1998yk} and confirmed by hand.
\item
Dimensional reduction is adopted, see Refs.~\cite{Hahn:2000kx,Kublbeck:1990xc,Hahn:1998yk}.
\end{itemize}

\subsubsection{Bubble and tadpole diagrams}

The two-point (``bubble'') diagrams involving bosons are given in Figs.~\ref{Fig:feynman-B1-B4}, \ref{Fig:feynman-B5-B10}
and \ref{Fig:feynman-Ba-Bc}. Their contributions to the amplitudes are given in table~\ref{Table:boson-B-amplitudes}.

\begin{table}[htb]
\caption{Amplitudes for bosonic bubble diagrams.}
\label{Table:boson-B-amplitudes}
\vspace*{2mm}
\resizebox{\columnwidth}{!}{
\begin{tabular}{|c|c|c|c|}
\hline
d & $K_\dd$ &  ${\cal F}_i^{\dd}$ &  Argument  \\
\hline
B-1& $\frac{g^2  s_{\text{W}}^2}{2 v c_{\text{W}}}$  & $q_iB_0$ & $\left(s_1,m_i^2,m_{{H^\pm}}^2\right)$  \\
\hline
B-2& $\frac{ g^2  s_{\text{W}}^2}{2 v^3 c_{\text{W}}}$  & $e_i\left(m_i^2-m_{{H^\pm}}^2\right)B_0$ & $\left(s_1,m_i^2,m_W^2\right)$  \\
\hline
B-3& $\frac{ g^4 s_{\text{W}}^2}{4 v c_{\text{W}}}$  & $e_iB_0$ & $\left(s_2,m_i^2,m_W^2\right)$  \\
\hline
B-4& $\frac{ g^4 s_{\text{W}}^2}{4 v c_{\text{W}}^3}$  & $e_iB_0$ & $\left(s_3,m_i^2,m_Z^2\right)$  \\
\hline
B-5& $\frac{g^2  s_{\text{W}}^2}{2 v c_{\text{W}} \left(s_1-m_W^2\right)}$  & $q_i \left(m_i^2-m_{{H^\pm}}^2\right)B_0$ & $\left(s_1,m_i^2,m_{{H^\pm}}^2\right)$  \\
\hline
B-6& $\frac{ g^2  s_{\text{W}}^2}{2 v^3 c_{\text{W}} \left(s_1-m_W^2\right)}$  & $e_im_i^2 \left(m_i^2-m_{{H^\pm}}^2\right)B_0$ & $\left(s_1,m_i^2,m_W^2\right)$  \\
\hline
B-7& $\frac{ g^4 s_{\text{W}}^2}{4 v c_{\text{W}} \left(s_1-m_W^2\right)}$  & $e_i(\frac{m_W^2+m_i^2}{4}-\frac{s_1}{12}-2 B_{00}- s_1(\frac{1}{2} B_0- B_1+\frac{1}{2} B_{11}))$ & $\left(s_1,m_i^2,m_W^2\right)$  \\
\hline
B-8& $\frac{g^2  \left(s_3-s_2\right) c_{\text{W}}}{2 v \left(s_1-m_W^2\right)}$  & $q_i(B_0+2B_1)$ & $\left(s_1,m_i^2,m_{{H^\pm}}^2\right)$  \\
\hline
B-9& $\frac{ g^2 \left(s_3-s_2\right) c_{\text{W}} }{2 v^3 \left(s_1-m_W^2\right)}$  & $e_i\left(m_i^2-m_{{H^\pm}}^2\right)(B_0+2B_1)$ & $\left(s_1,m_i^2,m_W^2\right)$  \\
\hline
B-10& $\frac{ g^4 \left(s_3-s_2\right) c_{\text{W}}}{4 v \left(s_1-m_W^2\right)}$  & $e_i(B_0-B_1)$ & $\left(s_1,m_i^2,m_W^2\right)$  \\
\hline
\end{tabular} }
\end{table}

\begin{figure}[htb]
\begin{center}
\includegraphics[scale=0.5]{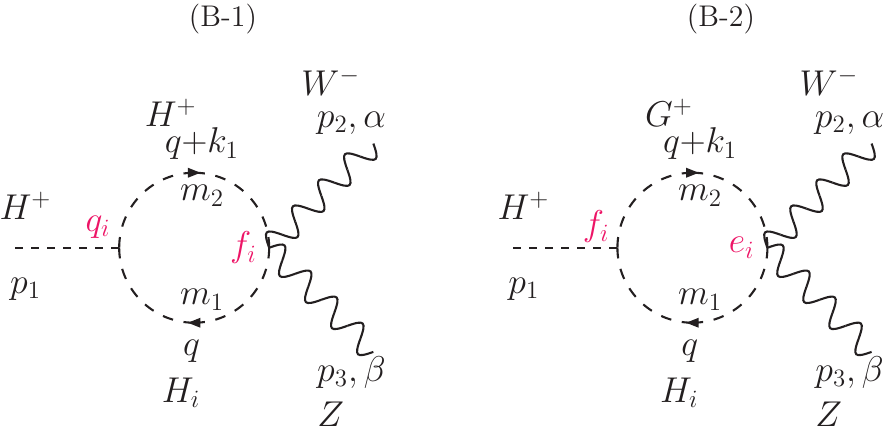} \hspace{8mm}
\includegraphics[scale=0.5]{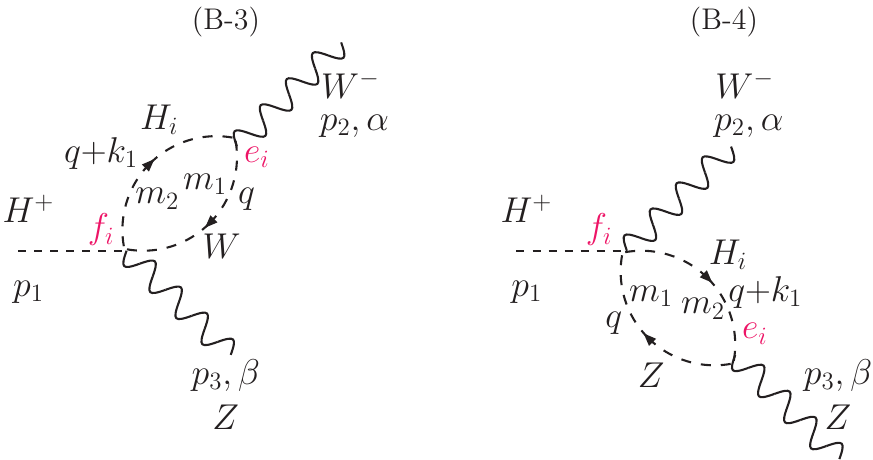}
\end{center}
\vspace*{-4mm}
\caption{Boson loops---parts~(B-1)--(B-4).}
\label{Fig:feynman-B1-B4}
\end{figure}

\begin{figure}[htb]
\begin{center}
\includegraphics[scale=0.5]{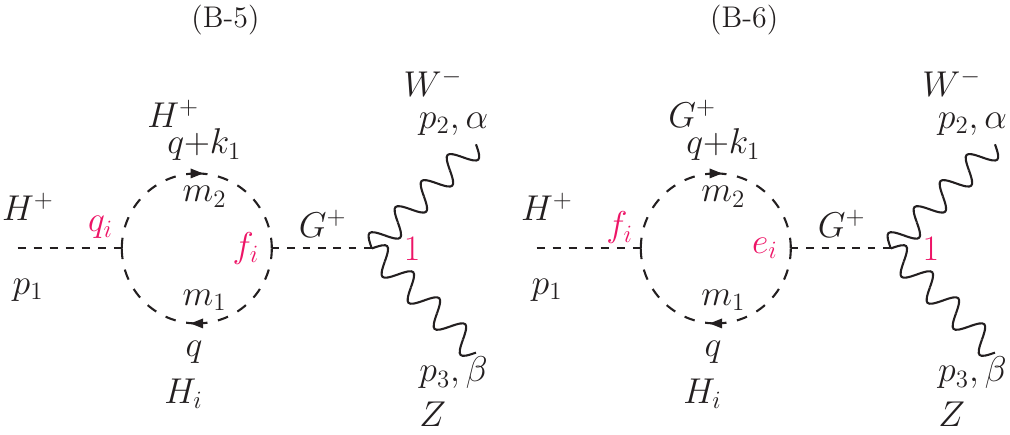} \
\includegraphics[scale=0.5]{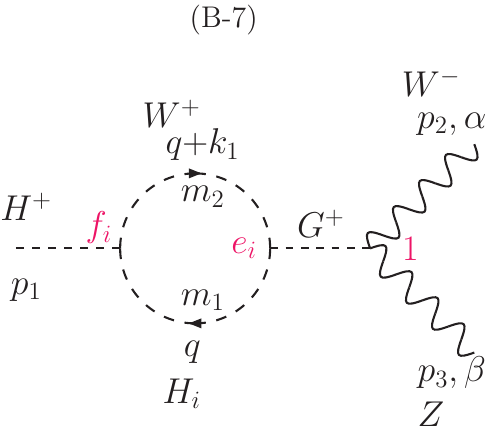} \\ \vspace{3mm}
\includegraphics[scale=0.5]{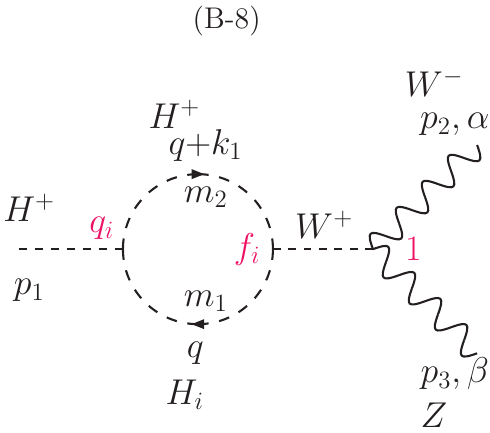} \
\includegraphics[scale=0.5]{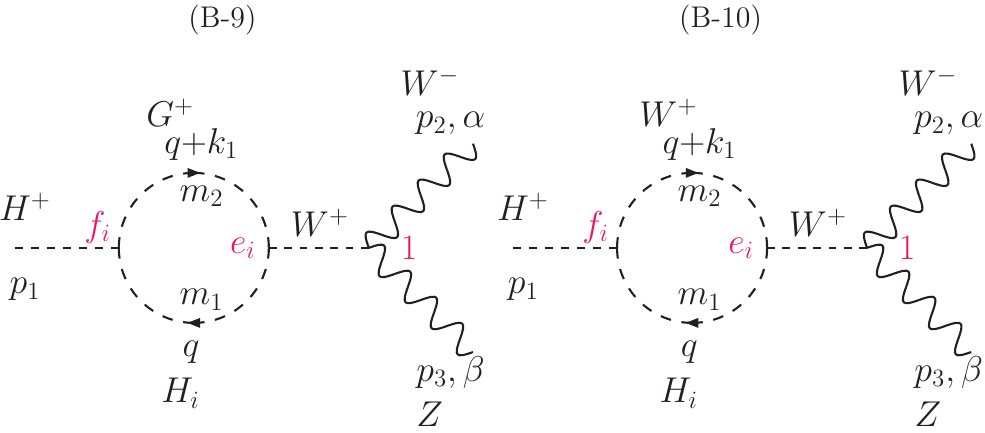}
\end{center}
\vspace*{-4mm}
\caption{Boson loops---parts~(B-5)--(B-10).}
\label{Fig:feynman-B5-B10}
\end{figure}

\begin{figure}[htb]
\begin{center}
\includegraphics[scale=0.5]{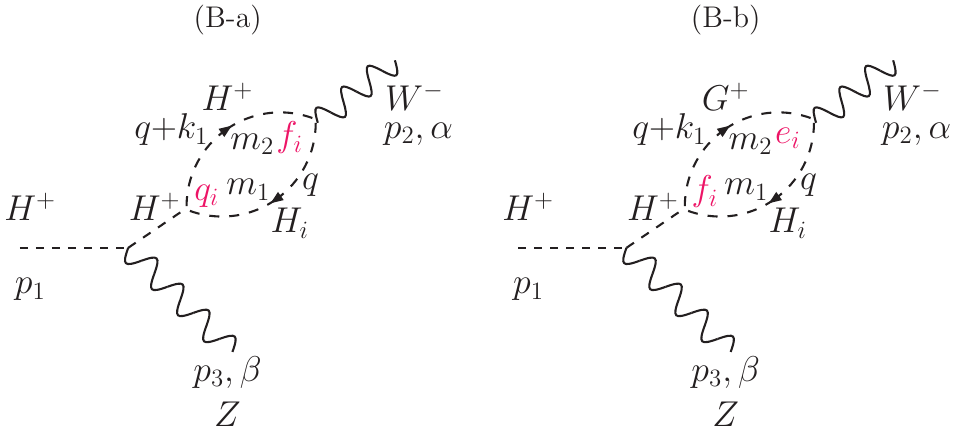} \
\includegraphics[scale=0.5]{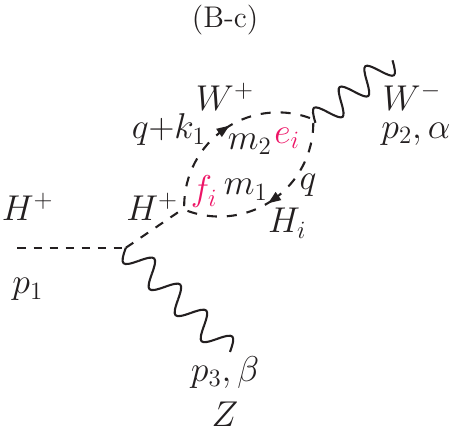}
\end{center}
\vspace*{-4mm}
\caption{Boson loops---parts~(B-a) to (B-c).}
\label{Fig:feynman-Ba-Bc}
\end{figure}

\begin{table}[htb]
	\caption{Amplitudes for bosonic tadpole diagrams}
	\label{Table:boson-T-amplitudes}
        \vspace*{2mm}
	\resizebox{\columnwidth}{!}{
		\begin{tabular}{|c|c|p{11cm}|}
			\hline
			d & $K_\dd$ &  \hspace*{5cm}${\cal F}_i^{\dd}$ \\
			\hline
			T-1& $\frac{g^2 s_{\text{W}}^2}{4 v^5 c_{\text{W}} \left(s_1-m_W^2\right)}$  & $2 e_i v^2 \left(m_i^2-m_{{H^\pm}}^2\right) A_0\left(m_i^2\right)
			+\displaystyle\sum_j(v^2(v^2-e_j^2)q_i-e_j^2e_im_i^2)A_0(m_j^2)$ \\
			\hline
			T-2& $\frac{ g^2  s_{\text{W}}^2}{4 v^3 c_{\text{W}} \left(s_1-m_W^2\right)}$  & $e_im_i^2A_0\left(m_Z^2\right)$ \\
			\hline
			T-3& $\frac{g^2  s_{\text{W}}^2}{v c_{\text{W}} \left(s_1-m_W^2\right)}$  & $q_iA_0\left(m_{{H^\pm}}^2\right)$ \\
			\hline
			T-4& $\frac{ g^2  s_{\text{W}}^2}{v^3 c_{\text{W}} \left(s_1-m_W^2\right)}$  & $e_im_i^2A_0\left(m_W^2\right)$ \\
			\hline
			T-5& $\frac{ g^4   s_{\text{W}}^2}{2 v c_{\text{W}}}$  & $\frac{\rho e_i}{m_i^2}(m_W^2-2 A_0\left(m_W^2\right))$ \\
			\hline
			T-6& $\frac{ g^4   s_{\text{W}}^2}{4  v c_{\text{W}}^3}$  & $\frac{\rho e_i}{m_i^2}(m_Z^2-2 A_0\left(m_Z^2\right))$ \\
			\hline
			T-7& $-\frac{ g^2   s_{\text{W}}^2}{2 v^3 c_{\text{W}}}$  & $\rho e_i A_0\left(m_W^2\right)$ \\
			\hline
			T-8& $-\frac{ g^2  s_{\text{W}}^2}{4 v^3 c_{\text{W}}}$  & $\rho e_i A_0\left(m_Z^2\right)$ \\
			\hline
			T-9& $-\frac{g^2  s_{\text{W}}^2}{2  v c_{\text{W}}}$  & $\frac{\rho  q_i}{m_i^2} A_0\left(m_{{H^\pm}}^2\right)$ \\
			\hline
			\multirow{4}*{T-10}& \multirow{4}*{$-\frac{g^2   s_{\text{W}}^2}{4  v^5 c_{\text{W}}}$}  & $\frac{\rho}{m_i^2}\bigl\{2 v^2  [2 e_i \left(m_i^2-m_{{H^\pm}}^2\right)+q_i v^2]A_0(m_i^2)$\par
			$\hspace*{0.8cm}+\displaystyle\sum_j[v^2(v^2-e_j^2)q_i-2v^2e_ie_jq_j-e_i(m_i^2-4m_{{H^\pm}}^2)e_j^2$\par
			$\hspace*{2cm}+2e_i(v^2-e_j^2)(m_j^2-m_{{H^\pm}}^2)]A_0(m_j^2)\bigr\}$ \\
			\hline
			T-11& $\frac{ g^4  s_{\text{W}}^2}{4  v c_{\text{W}}}$  & $\frac{\rho e_i}{m_i^2} A_0\left(m_W^2\right)$ \\
			\hline
			T-12& $\frac{ g^4   s_{\text{W}}^2}{8  v c_{\text{W}}^3}$  & $\frac{\rho e_i }{m_i^2}A_0\left(m_Z^2\right)$ \\
			\hline
		\end{tabular}}
\end{table}

Bubble diagrams on the $W$ line, like those shown in Fig.~\ref{Fig:feynman-Ba-Bc}, only contribute with terms proportional to $p_2^\alpha$, which we neglect in our approximation of on-shell $W$. Likewise, contributions from bubbles on the outgoing $Z$ would be proportional to $p_3^\beta$, which we also neglect.

\begin{figure}[htb]
\begin{center}
\includegraphics[scale=0.55]{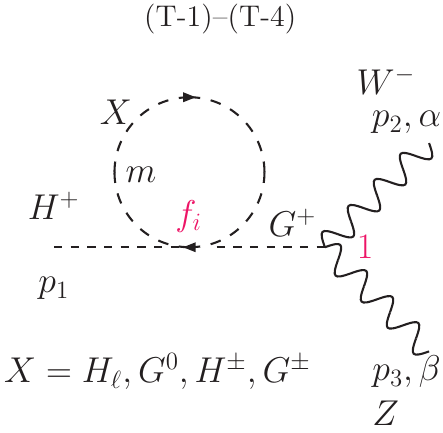}\ \hspace*{6mm}
\includegraphics[scale=0.55]{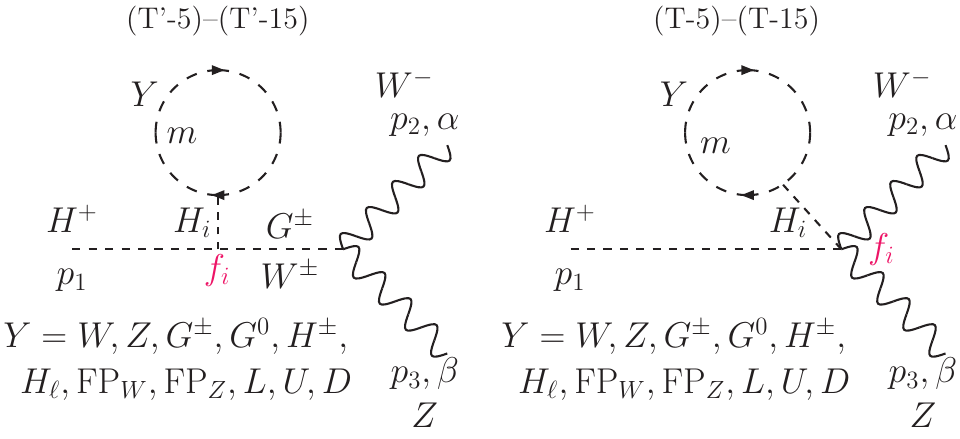}
\end{center}
\vspace*{-4mm}
\caption{Boson/fermion tadpole diagrams. Left: parts~(T-1)--(T-4). Centre and right: parts~(T-5)--(T-15).}
\label{Fig:feynman-bubble-in}
\end{figure}

For the diagrams in Fig.~\ref{Fig:feynman-bubble-in} we find the amplitudes given in the first four lines of table~\ref{Table:boson-T-amplitudes}, for loop particle $X\in\{H_\ell$, $G^0$, $H^\pm$, $Z\}$
(note that two identical fields do not yield a symmetry factor when forming a loop).

There are three sets of tadpole amplitudes arising from the diagrams in Fig.~\ref{Fig:feynman-bubble-in}. We denote the different fields running in the loop by $Y$. Furthermore, the diagram in the centre represents two cases ($W^\pm$ or $G^\pm$) for the field connecting the ``handle'' with the two external vector fields $W$ and $Z$. For each field $Y$, these two diagrams have a lot in common. For the diagram on the right, we find, with $Y\in \{W$, $Z$, $G^\pm$, $G^0$, $H^\pm$,  $H_\ell\}$, the results given in the lines (T-5) to (T-10) in table~\ref{Table:boson-T-amplitudes}.

There are also Faddeev--Popov ghost contributions. These are given in lines (T-11) and (T-12) of table~\ref{Table:boson-T-amplitudes}.

The contributions of the tadpoles in the centre of Fig.~\ref{Fig:feynman-bubble-in} with either $G^\pm$ or $W^\pm$ between the ``handle'' and the $WZ$ vertex, are related to those on the right-hand panel. In fact, we can include them in table~\ref{Table:boson-T-amplitudes} by adding their contributions to $A_i^{\text{(T-k)}\alpha\beta}$,
\begin{equation}
A_i^{\text{(T-k)}\alpha\beta}\to \rho A_i^{\text{(T-k)}\alpha\beta}, \quad k=5,\ldots 12,
\end{equation}
with
\begin{equation} \label{Eq:rho-def}
\rho=1+\frac{m_i^2-m_{H^\pm}^2}{s_1-m_{G^\pm}^2} + \frac{\cos^2\theta_W}{\sin^2\theta_W}\frac{s_3-s_2}{s_1-m_{G^\pm}^2},
\end{equation}
where ``1'' corresponds to the diagram on the right, and the two other terms correspond to those in the middle, with $G^\pm$ and $W^\pm$ respectively, between the ``handle'' and the $WZ$ vertex.

\subsubsection{Fermionic tadpole diagrams}
\label{sect:fermion-tadpoles}

In addition there are tadpole contributions, shown in Fig.~\ref{Fig:feynman-bubble-in} with a fermion loop. 
Due to the presence of the $f_i$ coupling, they are most conveniently evaluated with the ``bosonic'' amplitudes.

They follow the same structure as given in eqs. (\ref{famplitude}) and (\ref{conjugatefamplitude}). The results are given in table~\ref{Table:fermion-tadpole-amplitudes}. Here, the factor $\rho$ appearing in $K_\dd$ is given by Eq.~(\ref{Eq:rho-def}).
We note that the expression in the curly bracket originates from the Yukawa couplings of Eq.~(\ref{Eq:Scalar-fermion}), while the overall fermion mass factor comes from the trace.
Among these terms, the one involving the $t$-quark loop is obviously the dominant one.

\begin{table}[htb]
	\caption{Amplitudes for fermionic tadpole diagrams, where $s^2_\text{W}=\sin^2\theta_\text{W}$}
	\label{Table:fermion-tadpole-amplitudes}
	\begin{center}
		\begin{tabular}{|c|c|c|c|}
			\hline
			d & $K_\dd$ &  ${\cal F}_i^{\dd}$ \\
			\hline
			T-13& $\frac{g^2 \rho\,  s_{\text{W}}^2 }{2  v^3 c_{\text{W}}}$ & $\frac{ m_{l_k}\left\{\sqrt{2} v [(f_i{} \tilde{\rho }_{ll}^L{})^*+f_i \tilde{\rho }_{ll}^L]+4 e_i m_{l_k}\right\}}{m_i^2}A_0\left(m_{l_k}^2\right)$\\
			\hline
			T-14& $\frac{3 g^2 \rho\,  s_{\text{W}}^2 }{2  v^3 c_{\text{W}}}$ & $\frac{ m_{u_m}\left\{\sqrt{2} v [(f_i{} \tilde{\rho }_{mm}^U{})^*+f_i \tilde{\rho }_{mm}^U]+4 e_i m_{u_m}\right\}}{m_i^2}A_0\left(m_{u_m}^2\right)$\\
			\hline
			T-15& $\frac{3 g^2 \rho\,   s_{\text{W}}^2}{2  v^3 c_{\text{W}}}$ & $\frac{ m_{d_k}\left\{\sqrt{2} v [(f_i{} \tilde{\rho }_{kk}^D{})^*+ f_i  \tilde{\rho }_{kk}^D]+4 e_i m_{d_k}\right\}}{m_i^2}A_0\left(m_{d_k}^2\right)$\\
			\hline
				\end{tabular}
	\end{center}
\end{table}

Excluding the overall (complex) $f_i$ (or $f_i^\ast$, see Eqs.~(\ref{famplitude}) and (\ref{conjugatefamplitude})), these fermionic tadpole amplitudes are all real. This can be seen as follows: (1) the loop function $A_0(m^2)$ is real; (2), the curly bracket in table~\ref{Table:fermion-tadpole-amplitudes} contains terms with $f_i$, $f_i^\ast$ and $e_i$. The first two terms sum to a real quantity, whereas the third one (with $e_i$) is separately real. These terms depend on the sum of the phases of $f_i$ and $\tilde\rho$, denoted $\theta_\text{BF}$ for the reference case of $f_1$ and $\zeta_\tau$ and discussed in section~\ref{sect:Phases}.

We observe from Table~\ref{Table:fermion-tadpole-amplitudes} that when $\tilde\rho={\cal O}(1)$, the lighter fermions of mass $m$ contribute with terms of ${\cal O}(m/m_3)$, not ${\cal O}((m/m_3)^2)$, relative to the dominant ones, given by $m_3\in\{m_t,m_b,m_\tau\}$.

\subsection{Alignment limit}
\label{sect:alignment}

Alignment refers to the fact that the LHC experiments find a Higgs boson ``signal strength'' close to unity, i.e., the observed value is close to the SM prediction. The PDG values are $1.00\pm0.08$ for $WW$ and $1.02\pm0.08$ for the $ZZ$ Higgs decay channels. In the 2HDM framework, we can re-express alignment in terms of the coupling modifier, $\kappa=e_1/v$ of Eq.~(\ref{Eq:alignment}), with ``signal strength'' the square of the coupling, i.e., $(e_1/v)^2>1-0.08=0.92$ at the $1$-$\sigma$ level \cite{ParticleDataGroup:2024cfk}.

We shall see that, in the alignment limit and in the absence of fermion-loop contributions, there is no CP violation if hitherto unknown neutral scalar masses are chosen to have sufficiently high values.

Let us discuss what happens in the exact alignment limit, where (modulo an overall unphysical phase on the $f_i$) we have
\begin{subequations} \label{Eq:Alignment}
\begin{alignat}{3}
e_1&=v, &\quad e_2&=0, &\quad e_3&=0, \\
f_1&=0, &\quad f_2&=v, &\quad f_3&=-iv. 
\end{alignat}
\end{subequations}

Because of Eq.~(\ref{Eq:Alignment}) we only need to consider $F_{i}$ and $G_{i}$ for $i=2,3$, and furthermore only the parts that survive for $e_2=e_3=0$. This leaves only amplitudes that are proportional to $q_i$, representing coupling of a neutral scalar to a charged pair. The only triangle diagrams that contribute in this limit, are C-1 and C-2 (see Figs.~\ref{Fig:feynman-a} and \ref{Fig:feynman-b}, together with Eq.~(\ref{Eq:diagrams-a})). Likewise there are only three non-zero bubble (two-point) diagrams, B-1, B-5 and B-8 (see Figs.~\ref{Fig:feynman-B1-B4} and \ref{Fig:feynman-B5-B10}). Finally, the contributing tadpole diagrams are the bosonic T-1, T-3, T-9, T-10 and the fermionic $U$, $D$ and $L$ terms of table~\ref{Table:fermion-tadpole-amplitudes}.

For the ${\cal F}$ amplitudes, these particular diagrams are expressed in terms of loop integrals as follows:
\begin{subequations} \label{Eq:AL-terms}
\begin{alignat}{2}
&\text{C-1, C-2:} &\quad &C_{00}, \\
&\text{B-1, B-5, B-8:} &\quad &B_{0}, B_1,\\
&\text{T-1, T-3, etc.:} &\quad &A_{0}.
\end{alignat}
\end{subequations}
The $A_0$ function is real, while the others are in general complex. However, for these particular diagrams, due to the presence of a $q_i$ coupling, the $C_{00}$, $B_0$ and $B_1$ functions will all have one (or two) internal lines given by the $H^\pm$ field, it turns out that only diagram C-1 can possibly contain loop integrals with an imaginary part. This possibility stems from the $H_2H_3Z$ vertex. If the masses of the neutral scalars $H_2$ and $H_3$ are chosen such that $m_{2}+m_{3}>m_Z$, the resulting loop integrals will be real. Under these assumptions, it turns out that in the limit of alignment,
\begin{subequations}
\begin{align}
\tilde F_i&=-F_i^\ast, \quad i=2,3, \\
\tilde G_i&=-G_i^\ast, \quad i=2,3.
\end{align}
\end{subequations}
The absolute values are unchanged, there is no charge asymmetry.

\begin{figure}[htb]
\begin{center}
\includegraphics[scale=0.3]{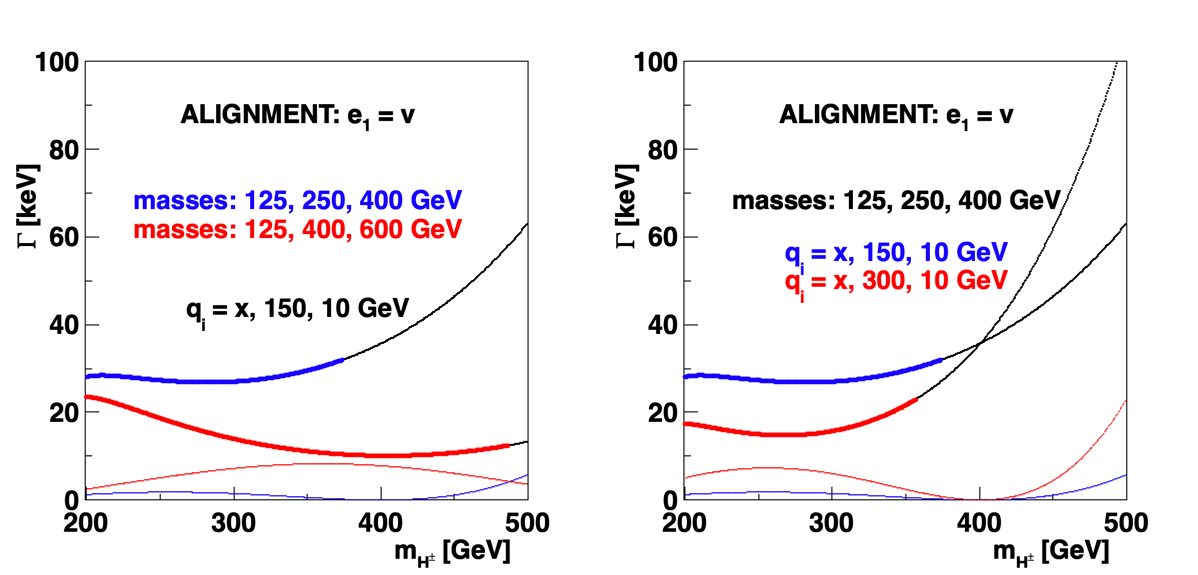}
\end{center}
\vspace*{-4mm}
\caption{Decay rates in the alignment limit at which $e_1=v$. Left: two sets of neutral masses $m_i$. Right: two sets of $q_i$ couplings involving the charged scalars. The coupling $q_1$ plays no role (denoted ``x'' in the figure), and the signs of $q_2$ and $q_3$ play no role. Solid upper curves include the fermionic tadpoles for $\tilde\rho^f=0.1$, the thin curves below do not.}
\label{Fig:AL-rate}
\end{figure}

Without the fermion amplitudes and for $m_{2}+m_{3}>m_Z$, there is no asymmetry in the alignment limit, but the couplings between the charged pair and the neutral scalars $H_i$ (denoted $q_i$) will lead to a finite decay rate, as illustrated in Fig.~\ref{Fig:AL-rate}. The case considered has $q_3$ small, and would have led to  ``small'' CP violation, if some loop integrals were complex. In this illustration, the contributions of fermion tadpoles are included, and (for the chosen parameters) do in fact dominate. In the left panel, we illustrate how the decay rate has some dependence on the masses of the neutral scalars. The fermionic tadpole amplitudes do not depend on the $q_i$ couplings, they are basically given by the $t$-quark mass and $\tilde\rho_{33}^U$, and obviously interfere with the amplitudes that are linear in $q_i$, as discussed above. In the right panel, we illustrate the dependence on $q_2$.

For the upper curves (including the tadpole contributions) we show in heavy colour (blue or red) the part that is consistent with the boundedness-from-below condition given in section~\ref{sect:Restrictions}.

Since in the alignment limit $f_1=0$, the decay rate does not depend on $q_1$ in this limit.
Also, since the amplitude is linear in $q_2$ and $q_3$, if we disregard the fermionic tadpole contributions, the decay rate is obviously invariant under a simultaneous change of their signs. Actually, the moduli of the amplitudes are invariant under individual sign changes of $q_2$ and $q_3$. In order to see this, let us define
\begin{equation} \label{Eq:F_i}
F_i =N_0\sum_\dd K_\dd {\cal F}_i^\dd,
\end{equation}
and similarly $G_i$, $\tilde F_i$ and $\tilde G_i$.

With $a$ and $b$ real, we have (see Eq.~(\ref{Eq:AL-terms}) and table~\ref{Table:boson-C-amplitudes})
\begin{align}
X_{23}\equiv F_2 f_2 + F_3 f_3
&=X_\text{tadpole}+N_0[(a q_2+ib q_3)v+(a q_3 -i b q_2)(-iv)] \nonumber \\
&=X_\text{tadpole}+N_0(a-b)v(q_2-iq_3).
\end{align}

Then, disregarding the tadpole part, and recalling that $N_0$ is purely imaginary, under sign changes we have
\begin{subequations}
\begin{alignat}{2}
q_2&\to-q_2:&\quad X_{23} &\to X_{23} ^\ast, \\
q_3&\to-q_3:&\quad X_{23}  &\to -X_{23} ^\ast,
\end{alignat}
\end{subequations}
and similarly for $G_i$, $\tilde F_i$ and $\tilde G_i$. Thus, in the absence of the tadpole part (i.e., for $\tilde\rho_{33}^f\to0$) and in the strict alignment limit, the moduli of ${\cal M}(H^\pm\to W^\pm Z)$ are proportional to $\sqrt{q_2^2+q_3^2}$, the decay rate is independent of the individual signs of $q_2$ and $q_3$.

As mentioned above, in the alignment limit, we only have contributions for $i\in\{2,3\}$. In this limit, if also the $\tilde\rho^f$ are real (or all have the same phase), the fermionic tadpole contribution for $i=3$ vanishes, since $f_3$ is purely imaginary.

As discussed elsewhere (see for example section~4.3 of Ref.~\cite{Grzadkowski:2014ada}), the 2HDM permits significant CP violation in the bosonic sector, also in the alignment limit. This would originate from the couplings of charged scalars to the neutral ones, other than the SM-like scalar. Such CP violation can be represented by the CP-odd basis-invariant $\Im J_{30}$ given in appendix~\ref{sect:CP-2HDM}, a quantity that is bilinear in the $q_i$ couplings. In the alignment limit we have $e_1=v$, but the product $q_2q_3$ could be substantial, represented by $\Im J_{30}$. The reason this invariant does not yield any charge asymmetry here, is that the relevant loop integrals are real if $m_{2}+m_{3}>m_Z$, as discussed above.

{
To summarise, if we adopt the notation that $m_1=125~\text{GeV}$ and $m_1< m_2 \leq m_3$, then obviously $m_2+m_3>m_Z$, and in the alignment limit the relevant (bosonic) loop integrals are real (no absorptive parts). Hence, excluding fermion contributions, there is in the alignment limit no CP violation in $H^\pm\to W^\pm Z$. However, as mentioned above, CP violation might still arise in other bosonic processes (see Ref.~\cite{Grzadkowski:2014ada}).
}
\subsection{Near-alignment decay rate and asymmetry}

As we move away from alignment, the amplitudes involving the $e_i$ couplings start contributing as follows: 
\begin{enumerate}
\item
the $f_1$ is now non-zero (since $e_1<v$), and 
\item
the $e_2$ and $e_3$ are also non-zero. 
\end{enumerate}
Thus, in addition to the fermionic tadpoles, and still without fermion loops, it is no longer only the couplings $q_i$ that play a role.  The decay rate and the corresponding asymmetry are shown in Fig.~\ref{Fig:nearAL-rate-asym} as functions of $m_{H^\pm}$, the charged-Higgs mass, for some choices of parameters. We show in heavy colour (blue or red) the low-$m_{H^\pm}$ part that is consistent with the boundedness-from-below conditions given in section~\ref{sect:Restrictions}.

\begin{figure}[htb]
\begin{center}
\includegraphics[scale=0.3]{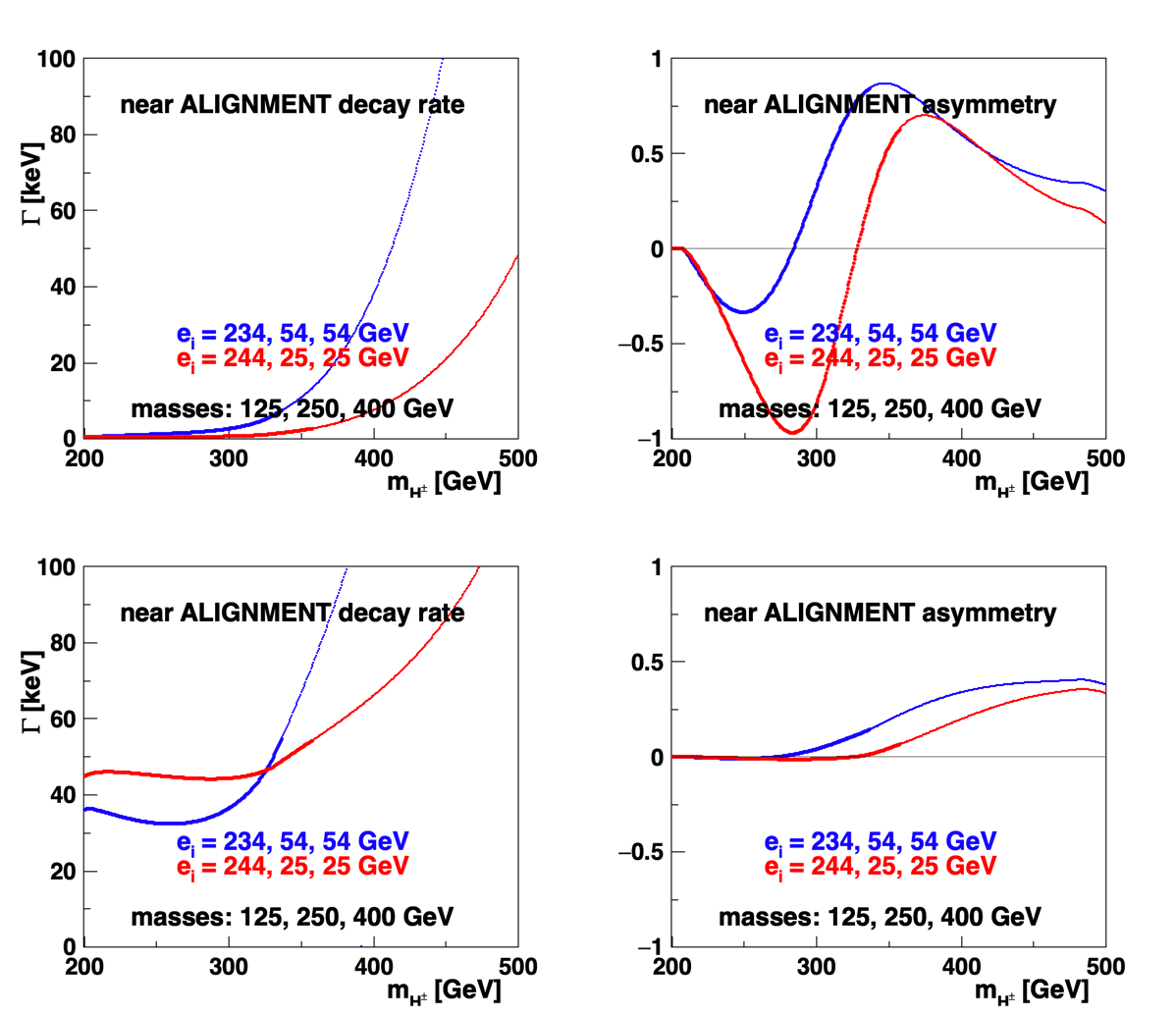}
\end{center}
\vspace*{-4mm}
\caption{Decay rate (left) and asymmetry (right) for some near-alignment cases. The upper panels are obtained {\it without} the fermion tadpole contributions, whereas the lower ones include them, for $\tilde\rho_{33}^U=\tilde\rho_{33}^D=\tilde\rho_{33}^L=0.1$ (real). The red curves represent a higher degree of alignment than the blue ones.}
\label{Fig:nearAL-rate-asym}
\end{figure}

Here, two phenomena are illustrated, 
\begin{itemize}
\item[(i)] the importance of the fermion tadpoles (lower {\it vs.}\ upper panels), and 
\item[(ii)] the importance of the degree of alignment (red vs blue). 
\end{itemize}
The two upper panels do not include the fermion tadpole contributions discussed in section~\ref{sect:fermions-tadpoles} (this corresponds to the limit $\tilde\rho^U=\tilde\rho^D=\tilde\rho^L\to0$), whereas the lower ones include them. 
{\color{blue}
Note that the  upper panels are based exclusively on bosonic amplitudes.
}

\paragraph{No fermionic tadpole contributions:}
At $H^\pm$ masses close to 350~GeV or higher, conflict with boundedness from below arises. At lower masses we can get a higher asymmetry, due to the condition for having complex loop amplitudes, mentioned in the Introduction, 
\begin{equation}
m_{H^\pm} > m_1+m_2,
\end{equation}
where $m_1$ and $m_2$ are the masses of the two loop fields that connect to the incoming $H^\pm$ field. The lowest possibility to have a complex amplitude is for these fields to be a $W^\pm$ (or a $G^\pm$) and an SM-like neutral Higgs boson, $m_{H^\pm}\gtrsim(80+125)~\text{GeV}$. However, when complying with boundedness from below, the asymmetry is somewhat irrelevant, since the decay rate, without the fermionic tadpole contribution, is negligible.

\paragraph{Including fermionic tadpoles:}
These tadpoles are seen to make a dominant contribution to the decay rate, as discussed in the context of Fig.~\ref{Fig:AL-rate}. 
Including the fermion tadpole contributions, the decay rate is significant already from the threshold, $m_{H^\pm}\geq m_W+m_Z$, whereas the asymmetry vanishes in the vicinity of the threshold (see the lower panels of Fig.~\ref{Fig:nearAL-rate-asym}). 
 
 \paragraph{Approach to Alignment:}
For a discussion of the approach to alignment, let us recall that alignment means $e_i= v$, for some state $i$, which we take to be the lightest one, i.e., $i=1$. In Fig.~\ref{Fig:nearAL-rate-asym} the cases studied have $e_1=0.95v$ (blue) and $e_1=0.99v$ (red).
In view of the constraint given by eq. (\ref{Eq:constraint-e_i}), we define
\begin{equation}
\Delta e^2 = v^2-e_1^2, \quad e_2^2+e_3^2\equiv\Delta e^2\ll v^2.
\end{equation}
In this limit, $e_2$ and $e_3$ will both be small. We parametrize the sharing, between $e_2$ and $e_3$, of the ``remainder'' $\Delta e^2$ by a parameter $r$,
\begin{equation} \label{Eq:def-r}
e_2=\pm_a\sqrt{r \Delta e^2}, \quad e_3=\pm_b\sqrt{(1-r) \Delta e^2}, \quad 0\le r \le 1.
\end{equation}
The two signs $\pm_a$ and $\pm_b$ are independent.
In Fig.~\ref{Fig:nearAL-rate-asym}, for both degrees of alignment, we take a symmetric sharing of the remainder, $r=0.5$, with $e_2$ and $e_3$ both positive.

\begin{figure}[htb]
\begin{center}
\includegraphics[scale=0.3]{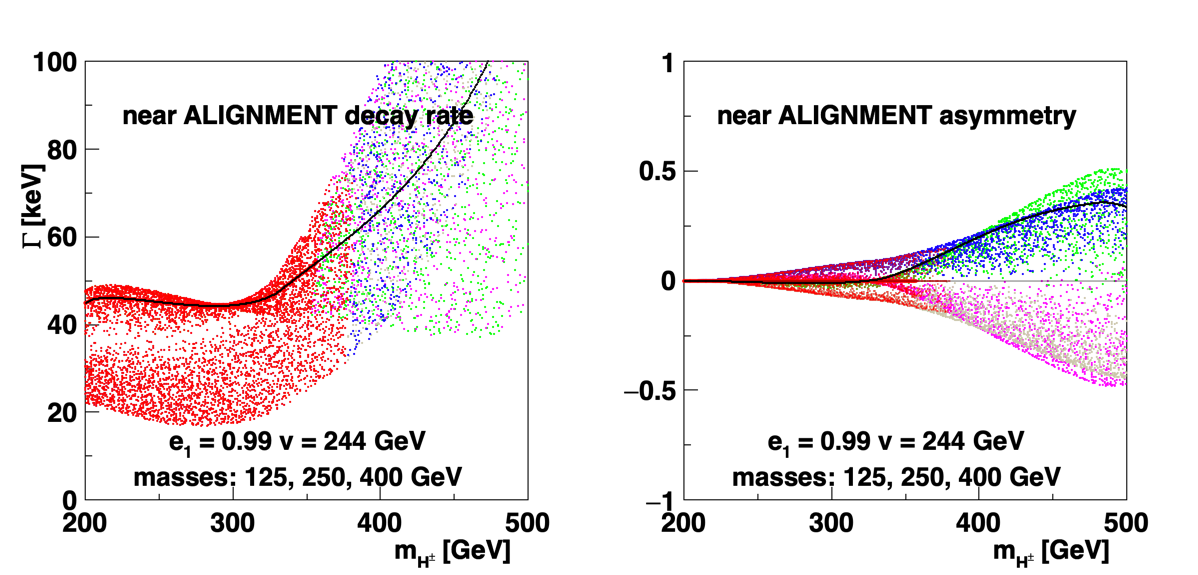}
\end{center}
\vspace*{-4mm}
\caption{Similar to the lower panels of Fig.~\ref{Fig:nearAL-rate-asym}, where the parameter $r$ of Eq.~(\ref{Eq:def-r}) was set to $0.5$, whereas here we scan over this parameter. This parameter controls the sharing of ``alignment deficit'' between $e_2$ and $e_3$. At high values of $m_{H^\pm}$ the decay rate and charge asymmetry are seen to be very sensitive to how the ``alignment deficit'' is shared on the two couplings $e_2$ and $e_3$.
See the text for identification of the colours.}
\label{Fig:nearAL-rate-asym-r}
\end{figure}

\begin{figure}[htb]
\begin{center}
\includegraphics[scale=0.27]{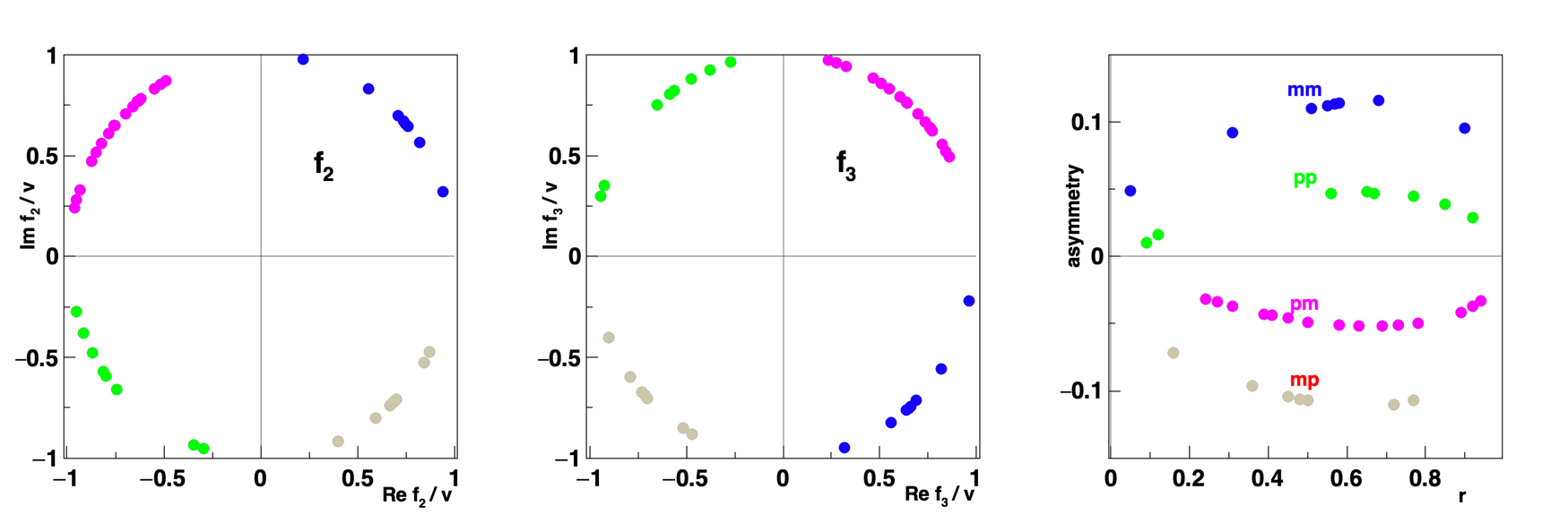}
\end{center}
\vspace*{-4mm}
\caption{Left and centre: Complex values of $f_2$ and $f_3$ for 40 random values of $r\in\{0,1\}$. Right: Corresponding asymmetries vs $r$, sorted into 4 sets, ``pp'' ($e_2>0$, $e_3>0$), ``pm'' ($e_2>0$, $e_3<0$), ``mp'' ($e_2<0$, $e_3>0$), ``mm" ($e_2<0$, $e_3<0$). The mass of $H^\pm$ is here 350~GeV.}
\label{Fig:couplings-vs-r}
\end{figure}

Next, in Fig.~\ref{Fig:nearAL-rate-asym-r} we illustrate the sensitivities that the decay rate and the asymmetry have to the parameter $r$ of Eq.~(\ref{Eq:def-r}) that quantifies the sharing of ``alignment deficit'' on the two couplings $e_2$ and $e_3$ and their signs. In this figure, we have scanned over $r$ for each point in $m_{H^\pm}$, allowing for both signs of $e_2$ and $e_3$, at the higher degree of alignment considered in Fig.~\ref{Fig:nearAL-rate-asym} (shown in red in the lower panels of that figure, for $e_1=0.99v=244~\text{GeV}$), keeping the other parameters fixed. 

The resulting values for the decay rate and asymmetry are shown as thin dots, together with the solid curve corresponding to $r=0.5$ and both $e_2$ and $e_3$ positive. Points that satisfy the boundedness-from-below conditions discussed in section~\ref{sect:Restrictions} are shown in red. In the upper range of $m_{H^\pm}$, we show in blue points where $e_2$ and $e_3$ are both negative, in violet (plotted on top of blue points) where $e_2>0$ and $e_3<0$, in brown where $e_2<0$ and $e_3>0$, and in green (plotted on top of the others) where both are positive.

One may conclude from this illustration that, even with strong alignment, there is considerable variation associated with both the decay rate and the asymmetry.

In order to better understand this spread in values for different combinations of $e_2$ and $e_3$ we consider in Fig.~\ref{Fig:couplings-vs-r} a scan over 40 values of $r$, with random signs for $e_2$ and $e_3$, correlating the resulting asymmetry at $m_{H^\pm}=350~\text{GeV}$ with the signs of $e_2$ and $e_3$. As was illustrated in Fig.~\ref{Fig:couplings-e-f}, these signs correlate with those of the $f_i$, which are more directly related to the asymmetry.  The colour coding is the same as in Fig.~\ref{Fig:nearAL-rate-asym-r}. 

While Eq.~(\ref{Eq:def-r}) shows that the moduli of $e_2$ and $e_3$ are interchanged under $r\leftrightarrow (1-r)$, the resulting asymmetries in the right panel of Fig.~\ref{Fig:couplings-vs-r} do not display this symmetry, since the masses $m_2$ and $m_3$ are different.
For the considered parameters, the asymmetry is positive when $e_2e_3>0$ and negative when $e_2e_3<0$.

The right panel of Fig.~\ref{Fig:nearAL-rate-asym-r} exhibits lines of higher density of scan points. These can be associated with the extrema seen in the parabola-like shapes in the right panel of Fig.~\ref{Fig:couplings-vs-r}.

\subsection{Mass degeneracy}
Kanemura and Mura do not report any asymmetry (no CP violation) in the purely bosonic sector. In the present study, we find such an asymmetry {\it away from alignment}. In their numerical studies, in order to suppress loop contributions to the electroweak precision observables $S$ and $T$, they adopt either of the following degeneracies, $m_{H^\pm}=m_3$ or $m_{H^\pm}=m_2$. The question then arises: does this degeneracy cause the loop integrals to become real, and thus prevent CP violation? 

For the case $m_{H^\pm}=m_3$, the amplitudes corresponding to an internal $H_3$ line like C-4 and C-5 are indeed real (the overall ``$i$'' from the rotation to the Euclidean space plays no role in this discussion). However, diagrams with $H_3$ are accompanied by diagrams with $H_1$ and $H_2$, these will in general be complex, since $m_1<m_2<m_3=m_{H^\pm}$. Similarly, for the $m_{H^\pm}=m_2$ degeneracy, with $m_1<m_2=m_{H^\pm}<m_3$, the loop diagrams with $H_1$ would be complex and generate an asymmetry. Thus, a mass degeneracy does not remove the CP violation in the purely bosonic sector\footnote{We do not consider $m_{H^\pm}=m_1$.}.

\section{Fermion contributions}
\label{sect:fermions}

The boson loop integrations involved up to two momentum-of-integration factors ($q^\alpha$ and/or $q^\beta$) in the numerator. This is also the case with the fermion loops, but in addition, a new feature emerges: there are terms where these two momenta are contracted, yielding $g^{\alpha\beta} g_{\mu\nu} q^\mu q^\nu$ or $g^{\alpha\beta} q^2$. We treat these integrals in dimensional reduction, they yield a contribution to the loop integral that is independent of the kinematics (energies and masses) \cite{Hahn:1998yk}.

\subsection{Quark ($tb$) loops}

There will be contributions to the decay amplitude from quark loops, some of which are depicted in Fig.~\ref{Fig:feynman-f}.  
The fermionic tadpoles were already included in the previous discussion.

\begin{figure}[htb]
\begin{center}
\includegraphics[scale=0.55]{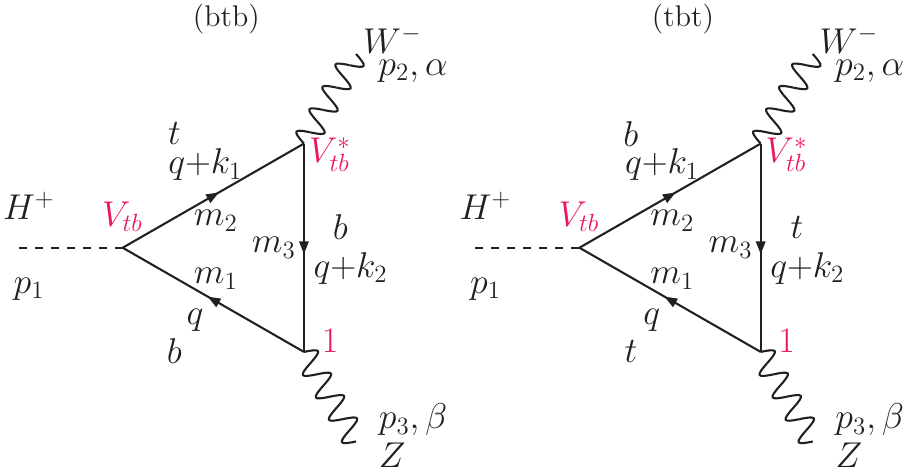}
\includegraphics[scale=0.60]{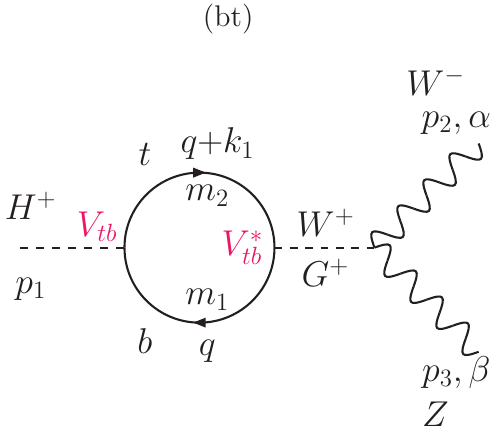}
\end{center}
\vspace*{-4mm}
\caption{Quark loops. The arrows indicate momentum flow, not charge flow.}
\label{Fig:feynman-f}
\end{figure}

Let us parametrize the $F$-, $G$- and $H$-amplitudes of the triangular quark diagrams by
\begin{subequations} \label{Eq:quark-decomposition}
\begin{eqnarray}
	F^\dd&=&N_0\,K_\dd\, V^\ast_{mk} \left([(\tilde\rho^U)^\dagger V]_{mk}\,{\cal F}_{mk}^{\dd,U} m_{u_m} +[V(\tilde\rho^D)^\dagger]_{mk}\, {\cal F}_{mk}^{\dd,D} m_{d_k}\right),\\
	G^\dd&=&N_0\, K_\dd\, V^\ast_{mk} \left([(\tilde\rho^U)^\dagger V]_{mk}\,{\cal G}_{mk}^{\dd,U}m_{u_m}+[V(\tilde\rho^D)^\dagger]_{mk}\, {\cal G}_{mk}^{\dd,D}m_{d_k} \right),\\
	H^\dd&=&iN_0\, K_\dd\, V^\ast_{mk} \left([(\tilde\rho^U)^\dagger V]_{mk}\,{\cal H}_{mk}^{\dd,U}m_{u_m}+[V(\tilde\rho^D)^\dagger]_{mk}\, {\cal H}_{mk}^{\dd,D} m_{d_k}\right).
\end{eqnarray}
\end{subequations}
Here, the index `d' refers to a set of Feynman diagrams, and the indices $m$ and $k$ are quark generations to be summed over, $m=1,2,3\, (u,c,t)$ and $k=1,2,3\, (d,s,b)$. (In practice, because of the strong mass hierarchy in the fermion sector, the dominant contributions come from $m=3$ and $k=3$.) We do not specify the arguments of the loop integrals, they will always be Passarino--Veltman functions $C_X\equiv C_X(s_1,s_2,s_3,m_1^2,m_2^2,m_3^2)$, where $s_1$, $s_2$ and $s_3$ are given by Eq.~(\ref{Eq:s_i}) whereas $m_1,m_2,m_3$ must be identified for each diagram by inspecting which fermions run in the loop. Furthermore, $V_{mk}$ is the CKM matrix element. The amplitudes of the corresponding charge-conjugate processes are given by
\begin{subequations}
\begin{eqnarray}
	\tilde{F}^\dd&=&N_0\,K_\dd^*\, V_{mk} \left([V^\dagger \tilde\rho^U]_{km}\,{\cal F}_{mk}^{\dd,U} m_{u_m} +[\tilde\rho^DV^\dagger]_{km}\, {\cal F}_{mk}^{\dd,D} m_{d_k}\right),\\
	\tilde{G}^\dd&=&N_0\, K_\dd^*\, V_{mk}\left([V^\dagger \tilde\rho^U]_{km}\,{\cal G}_{mk}^{\dd,U}m_{u_m}+[\tilde\rho^DV^\dagger]_{km}\, {\cal G}_{mk}^{\dd,D}m_{d_k} \right),\\
	\tilde{H}^\dd&=&-iN_0\, K_\dd^*\, V_{mk}\left([V^\dagger \tilde\rho^U]_{km}\,{\cal H}_{mk}^{\dd,U}m_{u_m}+[\tilde\rho^DV^\dagger]_{km}\, {\cal H}_{mk}^{\dd,D} m_{d_k}\right),
\end{eqnarray}
\end{subequations}
where $U$ and $D$ refer to up- and down-type quarks.

With the decomposition of Eq.~(\ref{Eq:quark-decomposition}), and the charged-Higgs and Goldstone Yukawa couplings given by Eq.~(\ref{Eq:Yukawas}), we obtain the amplitudes given in tables~\ref{Table:quarks-u-amplitudes}--\ref{Table:q-bubble-amplitudes}. Notice that in tables~\ref{Table:quarks-u-amplitudes} to \ref{Table:quarks-H-amplitudes} the $K_\dd$ factors are the same and therefore, are not repeated.
\begin{table}[htb]
	\caption{Amplitudes for quark triangle diagrams}
	\label{Table:quarks-u-amplitudes}
	\vspace*{2mm}
	\resizebox{\columnwidth}{!}{
		\begin{tabular}{|c|c|p{13cm}|c|}
			\hline
			d & $K_\dd$ & \hspace*{5cm} ${\cal F}_{mk}^{\dd,U}$ \\
			\hline
			\multirow{3}*{$d_ku_md_k$ (C-9)}& \multirow{3}*{$\frac{g^2}{2 \sqrt{2} c_{\text{W}}}$ } & $4s_{\text{W}}^2 m_{D_k}^2 C_0 +\left(3 c_{\text{W}}^2+s_{\text{W}}^2\right)\left(4 C_{00}-1\right) $\par $+s_1 \left(3 c_{\text{W}}^2+s_{\text{W}}^2\right) \left(C_1+2 \left(C_{11}+C_{12}\right)\right)-s_2\left(3 c_{\text{W}}^2+s_{\text{W}}^2\right) \left(C_1+2 C_{12}\right) $\par $+s_3 \left(3 c_{\text{W}}^2+s_{\text{W}}^2\right) \left(C_1+2 \left(C_{12}+C_2+C_{22}\right)\right)$\\
			\hline
			\multirow{4}*{$u_md_ku_m$ (C-10)} & \multirow{4}*{$-\frac{g^2}{2 \sqrt{2} c_{\text{W}}}$} & $12 C_{00} c_{\text{W}}^2+28 C_{00} s_{\text{W}}^2-3$\par $+s_1 \left(3 c_{\text{W}}^2 \left(C_0+3 C_1+C_2\right)+s_{\text{W}}^2 \left(-C_0+5 C_1+3 C_2\right)+6 \left(C_{11}+C_{12}\right)\right)$\par $+s_2 \left(-3 c_{\text{W}}^2 \left(C_0+C_1+C_2\right)+s_{\text{W}}^2 \left(C_0+C_1-3 C_2\right)-6 C_{12}\right)$ \par $+s_3 \left(3 c_{\text{W}}^2 \left(C_0+C_1+3 C_2\right)-C_0 s_{\text{W}}^2-C_1 s_{\text{W}}^2+6 C_{12}+C_2 s_{\text{W}}^2+6 C_{22}\right)$\\
			\hline
		\end{tabular}}
\end{table}

\begin{table}[htb]
	\caption{Amplitudes for quark triangle diagrams}
	\label{Table:quarks-d-amplitudes}
	\vspace*{2mm}
	\resizebox{\columnwidth}{!}{
		\begin{tabular}{|c|p{13cm}|c|}
			\hline
			d & \hspace*{5cm} ${\cal F}_{mk}^{\dd,D}$ \\
			\hline
			\multirow{4}*{$d_ku_md_k$ (C-9)} & $3-4\left(3 c_{\text{W}}^2+5 s_{\text{W}}^2\right) C_{00} $\par $+s_1 \left(-3 c_{\text{W}}^2 \left(C_0+3 C_1+C_2\right)-C_0 s_{\text{W}}^2-7 C_1 s_{\text{W}}^2-6 C_{11}-6 C_{12}-3 C_2 s_{\text{W}}^2\right)$\par $+s_2 \left(3 c_{\text{W}}^2 \left(C_0+C_1+C_2\right)+s_{\text{W}}^2 \left(C_0+C_1+3 C_2\right)+6 C_{12}\right)$\par $+s_3 \left(-3 c_{\text{W}}^2 \left(C_0+C_1+3 C_2\right)-C_0 s_{\text{W}}^2-C_1 s_{\text{W}}^2-6 C_{12}-5 C_2 s_{\text{W}}^2-6 C_{22}\right)$\\
			\hline
			\multirow{3}*{$u_md_ku_m$ (C-10)} & $s_{\text{W}}^2 \left(-8 C_0 m_{U_m}^2+4 C_{00}-1\right)+\left(3-12 C_{00}\right) c_{\text{W}}^2$\par $-s_1 \left(3 c_{\text{W}}^2-s_{\text{W}}^2\right) \left(C_1+2 \left(C_{11}+C_{12}\right)\right)+s_2 \left(C_1+2 C_{12}\right) \left(3 c_{\text{W}}^2-s_{\text{W}}^2\right)$\par $-s_3 \left(3 c_{\text{W}}^2-s_{\text{W}}^2\right) \left(C_1+2 \left(C_{12}+C_2+C_{22}\right)\right)$ \\
			\hline
		\end{tabular}}
\end{table}

The Yukawa coupling $\tilde\rho$ is in general complex, and this phase is independent of the complex phases of the loop integrals. Schematically, we can re-write the one-loop fermionic amplitudes as\footnote{We include lepton loops in this discussion.}
\begin{subequations}
\begin{align}
{\cal M}(H^+\to W^+Z)&=\sum_{f=U,D,L}\sum_\dd B_f^\text{(d)}(\tilde\rho^f_{ii})^\ast, \\
{\cal M}(H^-\to W^-Z)&=\sum_{f=U,D,L}\sum_\dd \tilde B_f^\text{(d)}\tilde\rho^f_{ii},
\end{align}
\end{subequations}
where again we do not show the $W^\pm$ and $Z$ polarization indices.

Summing the ${\cal F}^\dd$, ${\cal G}^\dd$ and ${\cal H}^\dd$ over the different diagrams `d', together with the pre\-factors $K_\dd$ of Eq.~(\ref{Eq:quark-decomposition}), we may write
\begin{subequations} \label{Eq:cal-M}
\begin{align}
{\cal M}(H^+\to W^+Z)&=\sum_{f=U,D,L}{\cal B}_f (\tilde\rho^f_{ii})^\ast, \\
{\cal M}(H^-\to W^-Z)&=\sum_{f=U,D,L} \tilde {\cal B}_f\tilde\rho^f_{ii}.
\end{align}
\end{subequations}

Here, ${\cal B}_f$ and $\tilde {\cal B}_f$ have identical moduli (note that the $K_\dd$ are real), but different phases. Assuming that the phases of the $\tilde\rho_{ii}^f$ are not all the same, i.e., that they differ for some $f$, then the sum over $f$ will lead to different moduli of the amplitudes in Eq.~(\ref{Eq:cal-M}). This will lead to a charge asymmetry.

While the phases $\zeta_f$ of $\tilde\rho^f$ are unphysical, phase differences are not.
We see that, with $f\in\{U,D,L\}$ (also referred to as $f\in\{t,b,\tau\}$), the asymmetry depends on the relative phases of the different $\tilde\rho$'s:
\begin{equation}
\zeta_t\neq \zeta_b \quad\text{and/or}\quad
\zeta_b\neq \zeta_\tau \quad\text{and/or}\quad
\zeta_\tau\neq \zeta_t,
\end{equation}
specified only for the heaviest fermions in each sector.


\begin{table}[htb]
	\caption{${\cal G}_{mk}^{\dd,U}$ amplitudes for quark triangle diagrams}
	\label{Table:quarks-G-amplitudes}
	\vspace*{2mm}
	\resizebox{\columnwidth}{!}{
		\begin{tabular}{|c|p{7cm}|p{7cm}|}
			\hline
			d & \hspace*{2cm} ${\cal G}_{mk}^{\dd,U}$ & \hspace*{2cm} ${\cal G}_{mk}^{\dd,D}$\\
			\hline
			\multirow{2}*{$d_ku_md_k$ (C-9)}& \multirow{2}*{$-2 \left(3 c_{\text{W}}^2+s_{\text{W}}^2\right) \left(C_1+2 \left(C_{11}+C_{12}\right)\right)$}
			& $6 c_{\text{W}}^2 \left(C_0+3 C_1+2 C_{11}+2 C_{12}+C_2\right)$\par $+2 s_{\text{W}}^2 \left(C_0+3 C_1+2 C_{11}+2 C_{12}+3 C_2\right)$\\
			\hline
			\multirow{2}*{$u_md_ku_m$ (C-10)} & $2 s_{\text{W}}^2 \left(C_0+3 C_1+2 C_{11}+2 C_{12}-3 C_2\right)$\par $-6 c_{\text{W}}^2 \left(C_0+3 C_1+2 C_{11}+2 C_{12}+C_2\right)$ & \multirow{2}*{$2 \left(3 c_{\text{W}}^2-s_{\text{W}}^2\right) \left(C_1+2 \left(C_{11}+C_{12}\right)\right)$} \\
			\hline
		\end{tabular}}
\end{table}

\begin{table}[htb]
	\caption{${\cal H}_{mk}^{\dd,U}$ and ${\cal H}_{mk}^{\dd,D}$ amplitudes for quark triangle diagrams}
	\label{Table:quarks-H-amplitudes}
	\vspace*{2mm}
	\resizebox{\columnwidth}{!}{
		\begin{tabular}{|c|p{7cm}|p{7cm}|}
			\hline
			d & \hspace*{2cm} ${\cal H}_{mk}^{\dd,U}$ & \hspace*{2cm} ${\cal H}_{mk}^{\dd,D}$\\
			\hline
			$d_ku_md_k$ (C-9)& $-2 C_1 \left(3 c_{\text{W}}^2+s_{\text{W}}^2\right)$
			& $6 c_{\text{W}}^2 \left(C_0+C_1\right)+2 s_{\text{W}}^2 \left(C_0+C_1\right)+6 C_2$\\
			\hline
			\multirow{2}*{$u_md_ku_m$ (C-10)} & $6 c_{\text{W}}^2 \left(C_0+C_1+C_2\right)$\par $-2 s_{\text{W}}^2 \left(C_0+C_1-3 C_2\right)$ & 
			\multirow{2}*{$2 C_1 \left(s_{\text{W}}^2-3 c_{\text{W}}^2\right)$} \\
			\hline
		\end{tabular}}
\end{table}

\begin{table}[htb]
	\caption{Amplitudes for quark bubble diagrams}
	\label{Table:q-bubble-amplitudes}
	\vspace*{2mm}
	\resizebox{\columnwidth}{!}{
		\begin{tabular}{|c|c|p{6.3cm}|p{6.5cm}|}
			\hline
			d & $K_\dd$ & \hspace*{2cm} ${\cal F}_{mk}^{\dd,U}$ & \hspace*{2cm} ${\cal F}_{mk}^{\dd,D}$ \\
			\hline
			\multirow{4}*{$d_ku_m (G^+)$ (B-11)} & \multirow{4}*{$\frac{g^2 s_{\text{W}}^2}{\sqrt{2} c_{\text{W}} \left(m_W^2-s_1\right)}$}  & 
			$s_1 \left(6 B_1\left(s_1,m_{d_k}^2,m_{u_m}^2\right)\right.$\par $\left.\hspace*{5mm}+6 B_{11}\left(s_1,m_{d_k}^2,m_{u_m}^2\right)+1\right)$\par $-3 \left(m_{D_k}^2 \left(2 B_0\left(s_1,m_{d_k}^2,m_{u_m}^2\right)+1\right)\right.$\par $\left.\hspace*{5mm}-8 B_{00}\left(s_1,m_{d_k}^2,m_{u_m}^2\right)+m_{u_m}^2\right)$ & $s_1 \left(6 B_1\left(s_1,m_{d_k}^2,m_{u_m}^2\right)\right.$\par $\left.\hspace*{5mm}+6 B_{11}\left(s_1,m_{d_k}^2,m_{u_m}^2\right)+1\right)$\par $-3 \left(m_{U_m}^2 \left(2 B_0\left(s_1,m_{d_k}^2,m_{u_m}^2\right)+1\right)\right.$\par $\left.\hspace*{5mm}-8 B_{00}\left(s_1,m_{d_k}^2,m_{u_m}^2\right)+m_{d_k}^2\right)$\\
			\hline
			\multirow{2}*{$d_ku_m (W^+)$ (B-12)} & \multirow{2}*{$\frac{3 \sqrt{2} g^2 \left(s_2-s_3\right) c_{\text{W}}}{m_W^2-s_1}$} & 
			\multirow{2}*{$B_1\left(s_1,m_{d_k}^2,m_{u_m}^2\right)$} & $-B_0\left(s_1,m_{d_k}^2,m_{u_m}^2\right)$\par $-B_1\left(s_1,m_{d_k}^2,m_{u_m}^2\right)$ \\
			\hline
		\end{tabular} }
\end{table}

The triangle diagrams will contribute to all three amplitudes, $F$, $G$ and $H$, since the two vector particles are individually connected to the fermion loop. On the other hand, the ``bubble'' diagrams to the right in Fig.~\ref{Fig:feynman-f} only have a single connection to the two vector particles, and thus only couple with a Lorentz structure given by $g^{\alpha\beta}$, i.e., they contribute only to the $F$ amplitude.

Since the loop integrals involved in the ${\cal H}$ amplitudes are all real, as are the $K_\dd$, the $H_\dd$ and $\tilde H_\dd$ amplitudes do not contribute to the difference $\Gamma(H^+\to W^+ Z) - \Gamma(H^-\to W^- Z)$ {\it unless} $(\tilde\rho^U)_{km}$ and/or $(\tilde\rho^D)_{km}$ contain non-diagonal elements, i.e., for some $k\neq m$. But they will in any case impact the asymmetry, via the denominator in Eq.~(\ref{Eq:asymmetry-def}).

\subsection{Lepton ($\tau\nu$) loops}

There are also contributions from lepton loops. They have the same topologies as those for quarks, given in Fig.~\ref{Fig:feynman-f}.

Let us parametrize the $F$, $G$ and $H$ amplitudes of the triangular lepton diagrams by
\begin{subequations}
\begin{eqnarray}
	F^\dd&=&N_0\,K_\dd\,  \,{\cal F}_{\ell}^{\dd,L} m_\ell ,\\
	G^\dd&=&N_0\, K_\dd\, {\cal G}_{\ell}^{\dd,L}m_\ell ,\\
	H^\dd&=&iN_0\, K_\dd\, \,{\cal H}_{\ell}^{\dd,L}m_\ell.
\end{eqnarray}
\end{subequations}
Again, the index `d' refers to a set of Feynman diagrams, whereas $\ell$ labels the lepton generations to be summed over, $\ell=1,2,3\, (e,\mu,\tau)$. The arguments of the loop integrals will be $C_X\equiv C_X(s_1,s_2,s_3,m_1^2,m_2^2,m_3^2)$, where $m_1,m_2,m_3$ must be identified for each diagram set by comparing with the corresponding Feynman diagram. The amplitudes of the corresponding charge-conjugated processes are given by
\begin{subequations}
\begin{eqnarray}
	\tilde{F}^\dd&=&N_0\,K_\dd^*\,  \,{\cal F}_{\ell}^{\dd,L} m_\ell ,\\
	\tilde{G}^\dd&=&N_0\, K_\dd^*\, {\cal G}_{\ell}^{\dd,L}m_\ell ,\\
	\tilde{H}^\dd&=&-iN_0\, K_\dd^*\, \,{\cal H}_{\ell}^{\dd,L}m_\ell.
\end{eqnarray}
\end{subequations}

The decomposition of eq.~(\ref{Eq:expansion})  yields for the two triangle diagrams of Fig.~\ref{Fig:feynman-f} the results summarised in tables~\ref{Table:lepton-triangle-F-amplitudes} and \ref{Table:lepton-G-H-amplitudes} whereas the results for the ``bubble'' diagrams are given in table~\ref{Table:amplitudes}. Notice that in tables~\ref{Table:lepton-triangle-F-amplitudes} and \ref{Table:lepton-G-H-amplitudes} the $K_\dd$ is the same. 

\begin{table}[htb]
	\caption{${\cal F}$ amplitudes for lepton triangle diagrams}
	\label{Table:lepton-triangle-F-amplitudes}
	\vspace*{2mm}
	\resizebox{\columnwidth}{!}{
		\begin{tabular}{|c|c|p{13.5cm}|c|}
			\hline
			d & $K_\dd$ & \hspace*{5cm} ${\cal F}_{\ell}^{\dd,L}$ \\
			\hline
			\multirow{4}*{$l_k \nu_{l_k} l_k$ (C-11)} & \multirow{4}*{$-\frac{g^2}{2 \sqrt{2} c_{\text{W}}}$}  & $\left(4 C_{00}-1\right) c_{\text{W}}^2+\left(12 C_{00}-1\right) s_{\text{W}}^2$\par $+s_1 \left(C_0 c_{\text{2W}}+c_{\text{W}}^2 \left(3 C_1+2 C_{11}+2 C_{12}+C_2\right)+s_{\text{W}}^2 \left(C_1+2 C_{11}+2 C_{12}+C_2\right)\right)$\par $+s_2 \left(-C_0 c_{\text{2W}}-\left(c_{\text{W}}^2 \left(C_1+2 C_{12}+C_2\right)\right)+s_{\text{W}}^2 \left(C_1-2 C_{12}-C_2\right)\right)$\par $+s_3 \left(C_0 c_{\text{2W}}+c_{\text{W}}^2 \left(C_1+2 C_{12}+3 C_2+2 C_{22}\right)-s_{\text{W}}^2 \left(C_1-2 C_{12}+C_2-2 C_{22}\right)\right)$\\
			\hline
			\multirow{2}*{$\nu_{l_k} l_k \nu_{l_k}$ (C-12)} & \multirow{2}*{$\frac{g^2}{2 \sqrt{2} c_{\text{W}}}$} & $4 C_{00}-1+s_1 \left(C_1+2 \left(C_{11}+C_{12}\right)\right)$\par $+s_2 \left(-C_1-2 C_{12}\right)+s_3 \left(C_1+2 \left(C_{12}+C_2+C_{22}\right)\right)$ \\
			\hline
		\end{tabular}}
\end{table}

\begin{table}[htb]
	\caption{${\cal G}$ and ${\cal H}$ amplitudes for lepton triangle diagrams}
	\vspace*{2mm}
	\label{Table:lepton-G-H-amplitudes}
	\resizebox{\columnwidth}{!}{
		\begin{tabular}{|c|p{7cm}|p{7cm}|}
			\hline
			d & \hspace*{3cm} ${\cal G}_\ell^{\dd,D}$ & \hspace*{3cm} ${\cal H}_\ell^{\dd,D}$ \\
			\hline
			\multirow{2}*{$l_k \nu_{l_k} l_k$ (C-11)} & $2 s_{\text{W}}^2 \left(C_0+3 C_1+2 C_{11}+2 C_{12}-C_2\right)$\par $-2 c_{\text{W}}^2 \left(C_0+3 C_1+2 C_{11}+2 C_{12}+C_2\right)$ & $2 s_{\text{W}}^2 \left(C_0+C_1-C_2\right)$\par $-2 c_{\text{W}}^2 \left(C_0+C_1+C_2\right)$\\
			\hline
			$\nu_{l_k} l_k \nu_{l_k}$ (C-12) & $-2 \left(C_1+2 \left(C_{11}+C_{12}\right)\right)$ & $2 C_1$\\
			\hline
		\end{tabular}}
\end{table}

\begin{table}[h!]
	\caption{Amplitudes for lepton bubble diagrams}
	\vspace*{2mm}
	\label{Table:amplitudes}
	\resizebox{\columnwidth}{!}{
		\begin{tabular}{|c|c|p{12cm}|}
			\hline
			$d$ & $K_d$ & \hspace*{2cm} ${\cal F}_{\ell}^{d,L}$ \\
			\hline
			$l_k \nu_{l_k} (G^+)$ (B-13)& $\frac{g^2 s_{\text{W}}^2}{3 \sqrt{2} c_{\text{W}} \left(s_1-m_W^2\right)}$  & $-24 B_{00}\left(s_1,0,m_{l_k}^2\right)-s_1 \left(6 B_1\left(s_1,0,m_{l_k}^2\right)+6 B_{11}\left(s_1,0,m_{l_k}^2\right)+1\right)+3 m_{{l_k}}^2$ \\
			$l_k \nu_{l_k} (W^+)$ (B-14) & $\frac{\sqrt{2} g^2 \left(s_3-s_2\right) c_{\text{W}}}{s_1-m_W^2}$ & $B_1\left(s_1,0,m_{l_k}^2\right)$  \\
			\hline
		\end{tabular}}
\end{table}

\begin{figure}[htb]
\begin{center}
\includegraphics[scale=0.25]{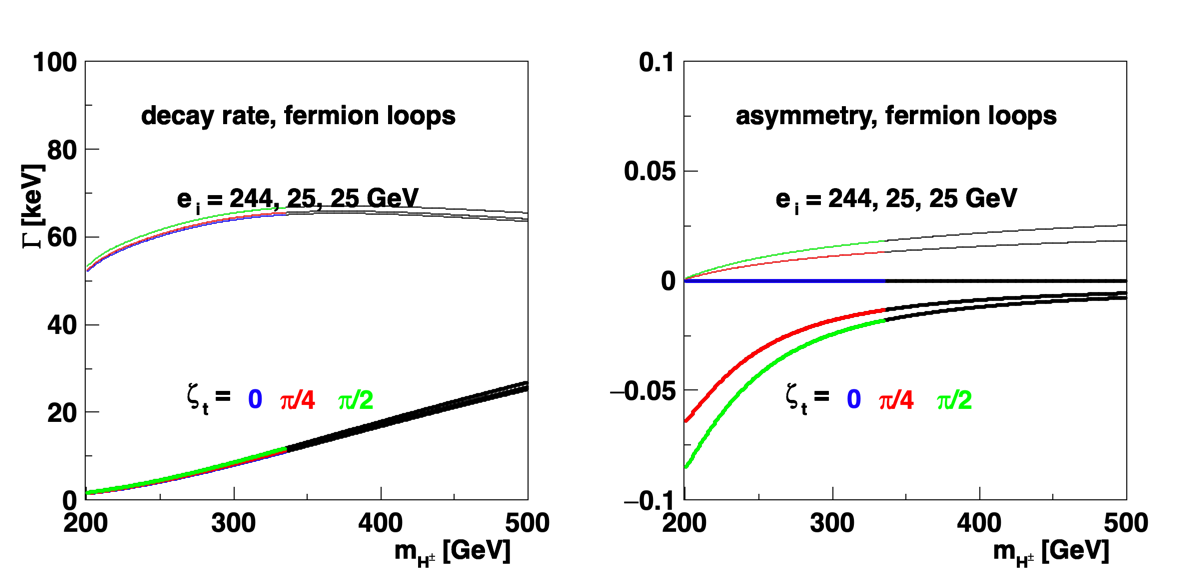}
\end{center}
\vspace*{-4mm}
\caption{Decay rate and asymmetry resulting from fermion loops only ($t$ and $b$ quarks and tau leptons with $\tilde\rho_{33}^X=0.1\times\exp(i\zeta_X)$, with $X=U, D, L$). The thicker curves include fermion tadpole diagrams, the thin curves (in the upper part) do not.}
\label{Fig:fermions-only}
\end{figure}

\subsection{Decay rate and asymmetry from fermionic diagrams}

Considering fermion loops only, the decay rate will obviously depend on the $\tilde\rho$ and their relative phases. In Type~I and Type~II 2HDMs, these are of the order $m/v$ (with $m$ being a fermion mass), but in the general case, these parameters are poorly constrained. If they all had the same phase, there would not be any charge asymmetry resulting from the fermionic sector (disregarding its interference with the bosonic sector). Also, in the limit of $m_t\gg m_b$, and the different $\tilde\rho^f$ being of the same order of magnitude, the amplitude proportional to $m_t$ has no counterpart against which to interfere, and the asymmetry would vanish. The same applies for the interference of the $t$-quark contributions with the leptonic contributions even though they do not run in the same loops as the quarks.

For a first illustration, we take the absolute values of $\tilde\rho$ to be 0.1, both for the $t$ quark and the $b$ quark (as well as for the less relevant $\tau$ lepton due to its lower mass). This particular value of 0.1 is the choice of Ref.~\cite{Kanemura:2024ezz}. 
Furthermore, a detailed study of anomalies in flavour physics was recently performed in the framework of the general 2HDM \cite{Athron:2024rir}. In fits to these anomalies, the authors found preferred parameter regions for $|\rho|={\cal O}(0.1)$.\footnote{These authors consider a general 2HDM, but impose CP conservation. They include factors $\cos(\beta-\alpha)$ and $\sin(\beta-\alpha)$ in their definitions of the parameter $\rho$. Thus, the comparison is not straightforward. See also Ref.~\cite{Crivellin:2013wna}} In the absence of interference with bosonic amplitudes, different phases are required for a non-zero asymmetry. 

In Fig.~\ref{Fig:fermions-only}, in addition to choosing $\zeta_\ell=0$, without loss of generality, we also choose $\zeta_b=0$ and three values of $\zeta_t$, namely 0, $\pi/4$ and $\pi/2$. This choice of $\zeta_b$ has little impact on our results.  Bosonic loops are not included.

\begin{figure}[htb]
\begin{center}
\vspace*{-4mm}
\includegraphics[scale=0.27]{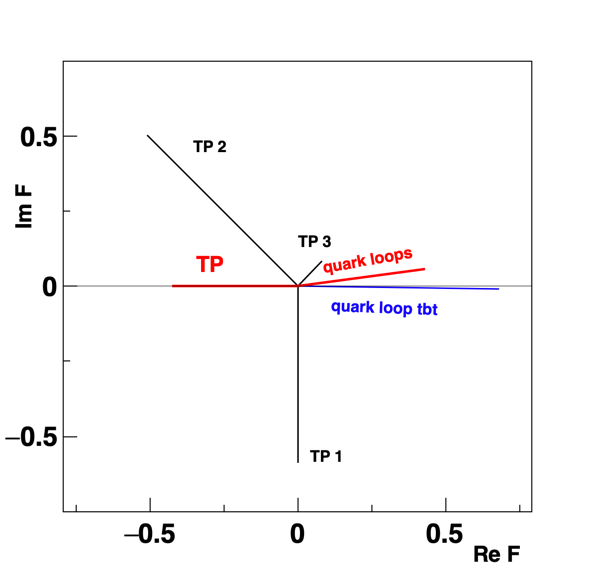}
\end{center}
\vspace*{-6mm}
\caption{Complex-plane vectors representing different contributions to the fermion amplitude ${\cal F}$ and the fermionic tadpole amplitudes. Here, we consider strong alignment, $e_1=0.99v$ and $\zeta_t=\pi/2$. The vectors labelled ``TP i'', with $i=1,2,3$, display the different contributions to the tadpole amplitude, proportional to $f_i$, with the red vector labelled ``TP'' showing the resulting (real) sum. In blue, we show the dominant fermion-loop term, namely the second row in table~\ref{Table:quarks-u-amplitudes}. In red (labelled ``quark loops''), we show the total ${\cal F}^U$ amplitude, summed over diagrams.}
\label{Fig:vectors}
\end{figure}

In contrast to the case of the bosonic diagrams, the fermionic ones develop an imaginary part, thus giving rise to a charge asymmetry immediately above the threshold for on-shell $H^\pm\to tb$ production.
Also, the fact that $m_t + m_b<m_{H_\text{SM}}+m_W$ means that, compared to the bosonic amplitudes, the diagrams involving fermions will have a faster rise at low values of $m_{H^\pm}$. This is easily seen by comparing Fig.~\ref{Fig:fermions-only} (for fermion loops only) with the upper panels of Fig.~\ref{Fig:nearAL-rate-asym} (bosonic loops only).

In Fig.~\ref{Fig:fermions-only}, the solid curves show the decay rate and asymmetry with the inclusion of the fermion tadpole contributions, whereas the thin curves respresent the pure fermion graphs only, i.e., excluding the tadpole contributions. Interestingly, there is a considerable cancellation between the fermion tadpole contributions and the pure fermionic contributions. For the considered parameters, this is due to destructive interference between the second (three-point) diagram in Fig.~\ref{Fig:feynman-f} and the $t$-quark tadpole diagrams in Fig.~\ref{Fig:feynman-bubble-in}. The coloured segments identify the range in $m_{H^\pm}$ where the potential is bounded from below, for the following choice of masses and couplings for the bosonic sector: $m_i=\{125, 200, 250\}~\text{GeV}$, $e_i=\{244, 25, 25\}~\text{GeV}$ and $q_i=\{100, 150, 10\}~\text{GeV}$.

Kanemura and Mura also explored values of $|\tilde\rho|$ as low as 0.01 \cite{Kanemura:2024ezz}. Adopting such low values of $|\tilde\rho|$, the rate shown in Fig.~\ref{Fig:fermions-only} (proportional to $|\tilde\rho|^2$) would be scaled down by a factor of $(0.01/0.1)^2=0.01$, leaving the asymmetry unchanged. We shall return to this issue in the next section. When the fermionic amplitude interferes with the bosonic one, there will be a regime where the dependence on $\tilde\rho$ is only linear, not quadratic.

The cancellation mentioned above, between the fermionic tadpoles and the fermionic triangle diagrams is illustrated in Fig.~\ref{Fig:vectors}, where we compare the sum of the tadpole contributions (which, as discussed, is real) with the contributions from the quark loops, focussing on those proportional to $m_t$. The parameters considered for this illustration, are those of Fig.~\ref{Fig:fermions-only}, with $\zeta_t=\pi/2$.

\section{Boson--fermion interference}
\label{sect:Boson-fermion}

\begin{figure}[htb]
\begin{center}
\includegraphics[scale=0.25]{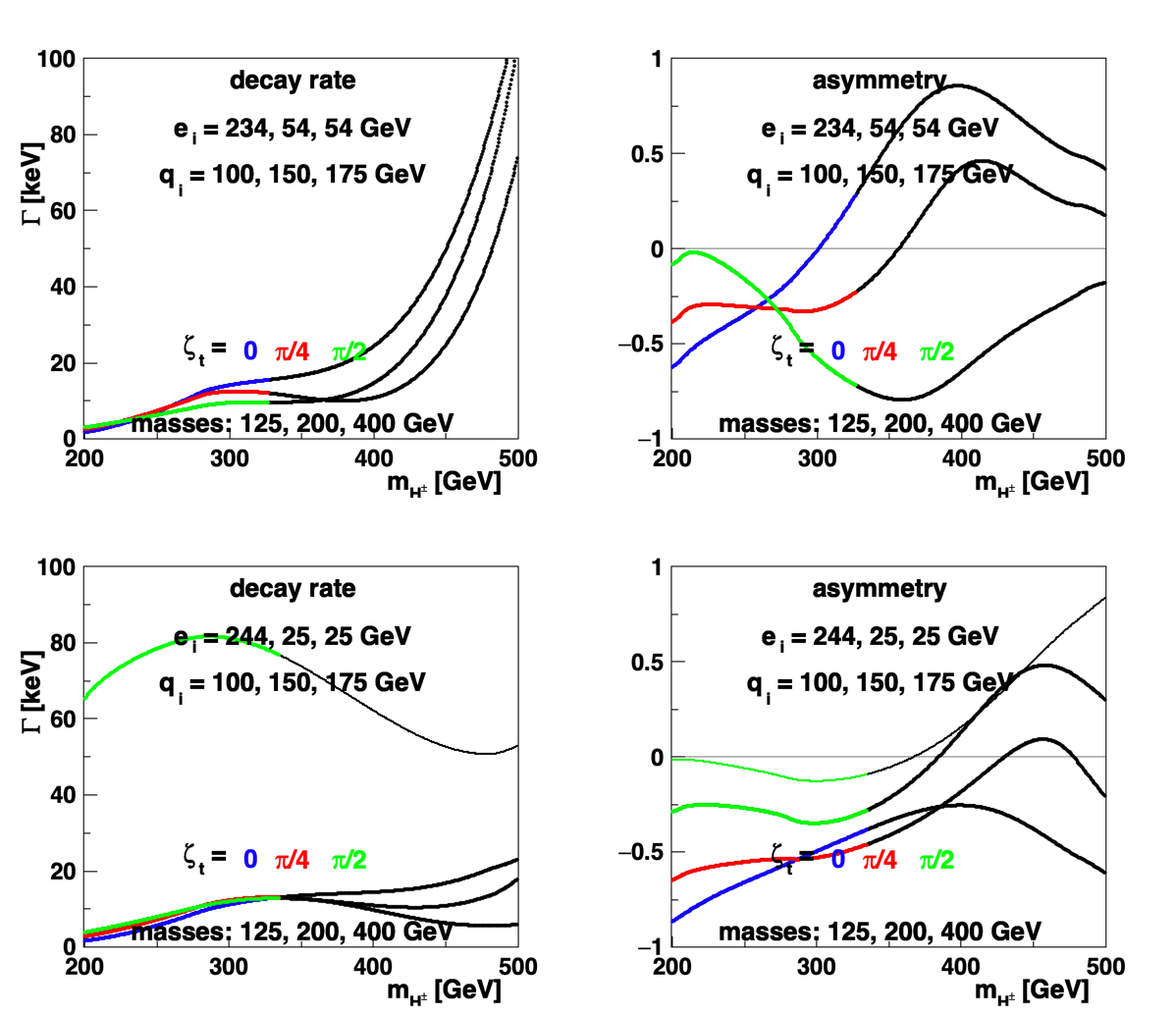}
\end{center}
\vspace*{-6mm}
\caption{Decay rate and asymmetry resulting from bosonic and fermionic loops ($t$ and $b$ quarks and tau leptons) for limits near alignment.
For the bosonic amplitudes, we consider two degrees of alignment, upper row: $e_1=0.95v$, lower row: $e_1=0.99v$. For the fermionic loops, we consider $\tilde\rho_{33}^X=0.1\times\exp(i\zeta_X)$, where $X$ is $t$, $b$, $\tau$ with $\zeta_b=\zeta_\tau=0$ and three values of $\zeta_t$. Thin, green curves in lower panels: no fermion tadpole contributions.}
\label{Fig:near-AL+fermions-a}
\end{figure}

Let us now include the contributions of both bosonic and fermionic loops. This opens up a complex multi-parameter scenario. As a first illustration, we show in Fig.~\ref{Fig:near-AL+fermions-a} the decay rate and asymmetry for the following parameter sets:
\begin{itemize}
\item
neutral-state masses: 125, 200 and 400~GeV;
\item
two degrees of alignment: upper panels: $e_1=0.95v$; lower panels: $e_1=0.99v$;
\item
couplings to charged pairs: $q_i=100, \ 150$ and 175~GeV;
\item
$|\tilde\rho^t|=|\tilde\rho^b|=|\tilde\rho^\tau|=0.1;$
\item
three phases $\zeta$ for the top Yukawa coupling: $\zeta_t=0$, $\pi/4$ and $\pi/2$; relative boson--fermion phase (see section~\ref{sect:Phases}) $\theta_\text{BF}=0$.
\end{itemize}

While the fermion tadpoles gave an important contribution to the results shown in Fig.~\ref{Fig:nearAL-rate-asym}, here they actually interfere destructively with the other amplitudes, and reduce the decay rate. We illustrate this in the two lower panels (for the higher degree of alignment), the thin green curves refer to the case of $\zeta_t=\pi/2$, like the heavier green curves, but now {\it without} the fermion tadpole contributions discussed in section~\ref{sect:fermion-tadpoles}. 

\begin{figure}[htb]
\begin{center}
\includegraphics[scale=0.25]{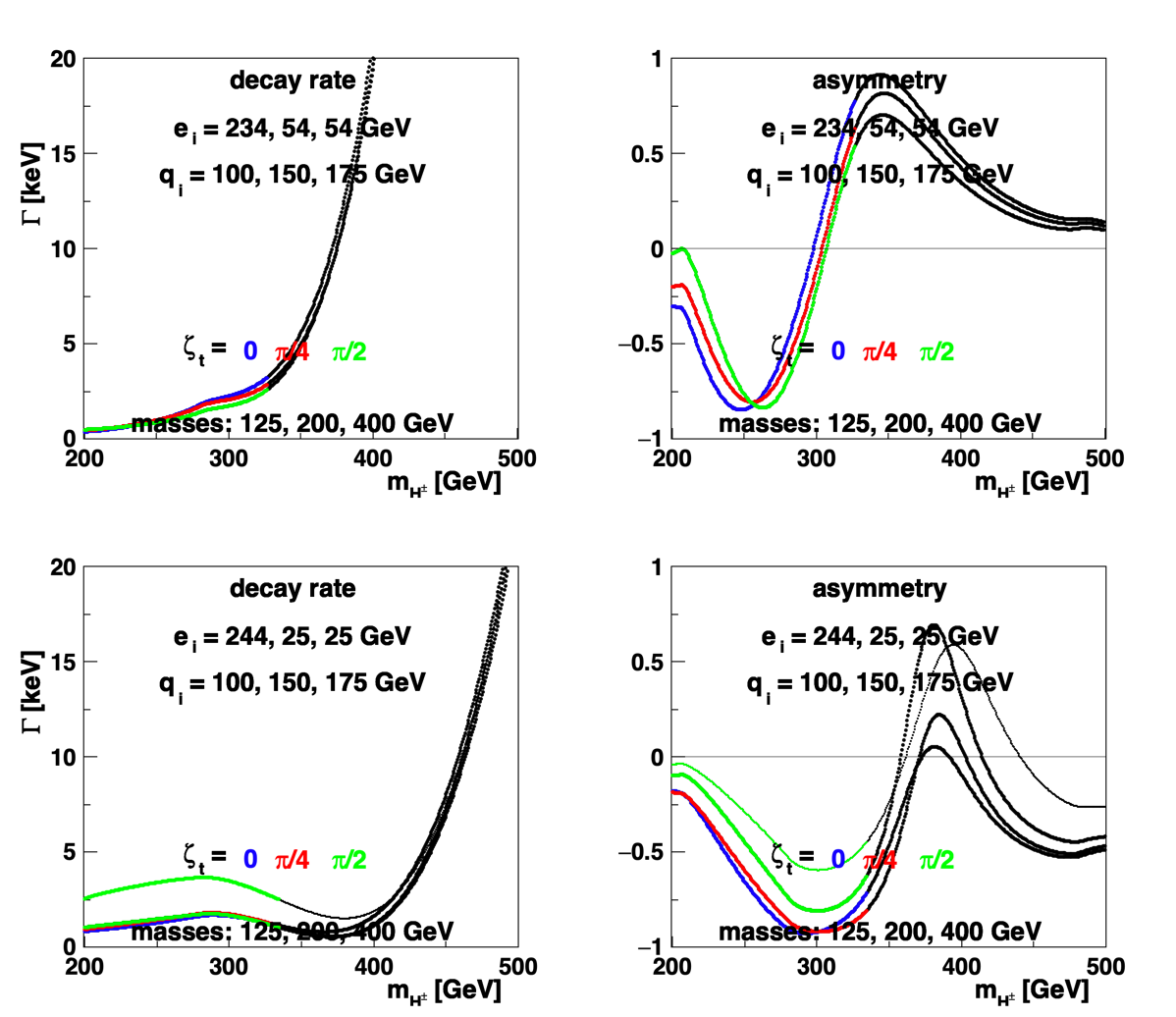}
\end{center}
\vspace*{-4mm}
\caption{Similar to Fig.~\ref{Fig:near-AL+fermions-a}, for $|\tilde\rho_{33}^U|=|\tilde\rho_{33}^D|=|\tilde\rho_{33}^L|=0.01$, with the same choice of $\zeta$'s. Note that the range in $\Gamma$ here is different.}
\label{Fig:near-AL+fermions-b}
\end{figure}

\begin{figure}[htb]
\begin{center}
\includegraphics[scale=0.25]{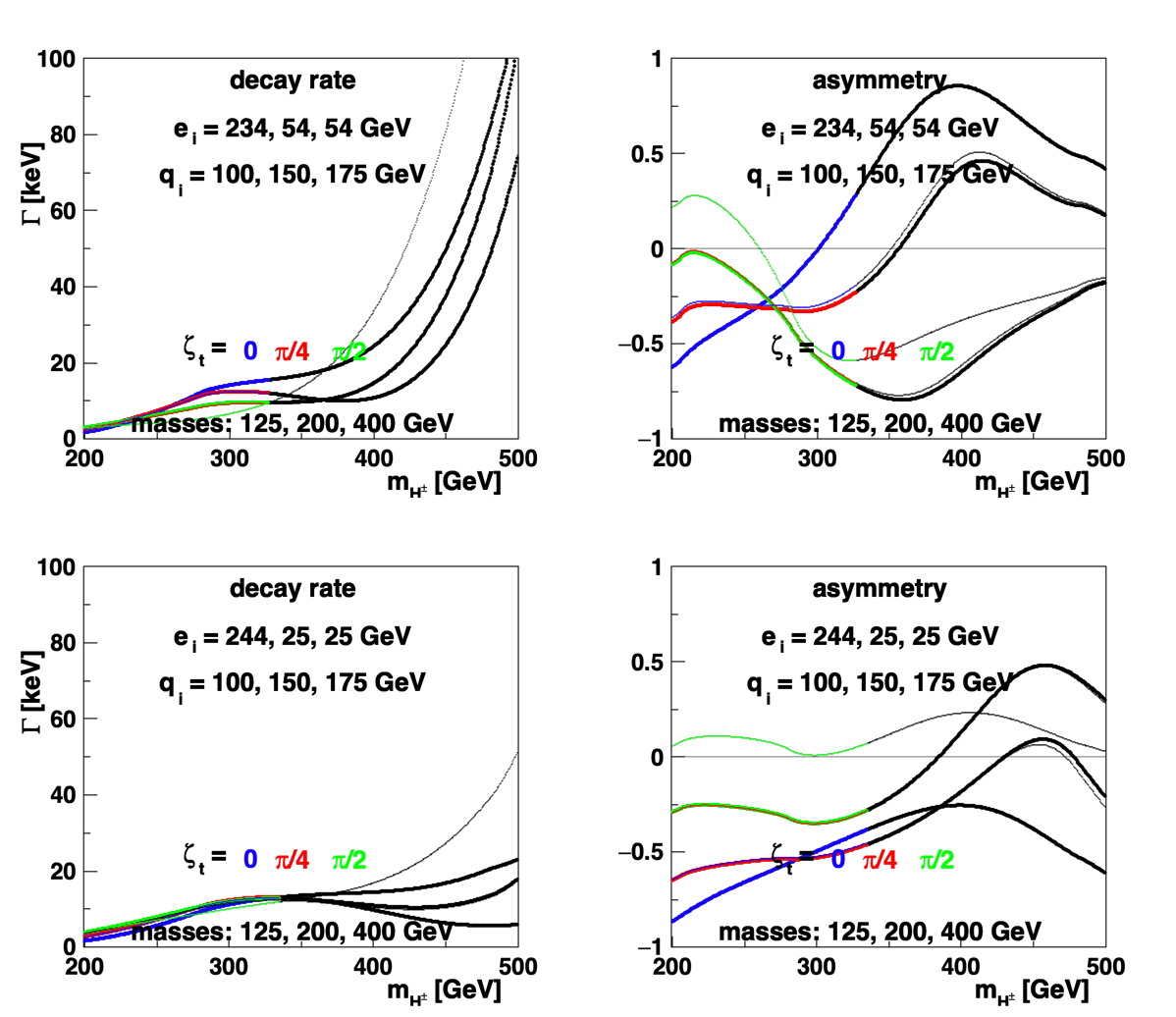}
\end{center}
\vspace*{-4mm}
\caption{Similar to Fig.~\ref{Fig:near-AL+fermions-a}, for $\theta_\text{BF}=0$ (heavy curves) and $\theta_\text{BF}=\pi/4$ (thin curves).}
\label{Fig:near-AL+fermions-c}
\end{figure}

Comparing the two cases of different degrees of alignment, we observe for low $H^\pm$ masses (where the boundedness-from-below conditions are satisfied) a quantitative, but no qualitative difference. This can be attributed, for the considered parameters, to the dominant contribution of the loops involving fermions, tadpoles and pure fermionic loops, as illustrated in the previous section. At higher $H^\pm$ masses, however, the case of less alignment (upper panels) leads to a significantly higher decay rate.

In order to illustrate the importance of $\tilde\rho$, in Fig.~\ref{Fig:near-AL+fermions-b} we have taken the $\tilde\rho$ Yukawa coupling parameters smaller by a factor of 0.1, i.e., here we take $|\tilde\rho^t|=|\tilde\rho^b|=|\tilde\rho^\tau|=0.01$. This brings the decay rate closer to that of the pure bosonic case, shown in the upper panels of Fig.~\ref{Fig:nearAL-rate-asym}. With such small Yukawa couplings, the sensitivity of the decay rate to the relative phases of the Yukawa couplings also becomes significantly reduced. The asymmetry has a considerable sensitivity to the relative phases $\zeta$, but experimental access may be difficult because of the small decay rate.

The above Figs.~\ref{Fig:near-AL+fermions-a} and \ref{Fig:near-AL+fermions-b} are obtained for a vanishing relative phase between the bosonic and fermionic sectors, $\theta_\text{BF}=0$ (see section~\ref{sect:Phases}). Next, we briefly discuss the impact of this parameter. In Fig.~\ref{Fig:near-AL+fermions-c} we show decay rates and asymmetries corresponding to those shown in Fig.~\ref{Fig:near-AL+fermions-a}, but now comparing $\theta_\text{BF}=0$ (heavy curves) with $\theta_\text{BF}=\pi/4$ (thin curves).

In the limit of $m_\tau,m_b\ll m_t$, the relative $\zeta$ phases in the fermionic sector play no role. Thus, the effect of a non-zero $\theta_\text{BF}$ is the same as that of a non-zero difference $\zeta_t-\zeta_\tau$. This is borne out in the figure: the thin blue curves (for $\theta_\text{BF}=\pi/4$ and $\zeta_t=0$) are seen to be close to the heavy red ones (for $\theta_\text{BF}=0$ and $\zeta_t=\pi/4$). Likewise, the thin red ones  (for $\theta_\text{BF}=\pi/4$ and $\zeta_t=\pi/4$) are seen to be close to the heavy green ones (for $\theta_\text{BF}=0$ and $\zeta_t=\pi/2$). The finite masses of the $b$ quark and the $\tau$ lepton lead to a small deviation from this simple limit. Thus, in the limit $m_b/m_t\ll1$ and $m_\tau/m_t\ll1$, we would have an effective phase,
\begin{equation}
\zeta_t^\text{eff}=\theta_\text{BF}+\zeta_t.
\end{equation}

\begin{figure}[htb]
\begin{center}
\includegraphics[scale=0.25]{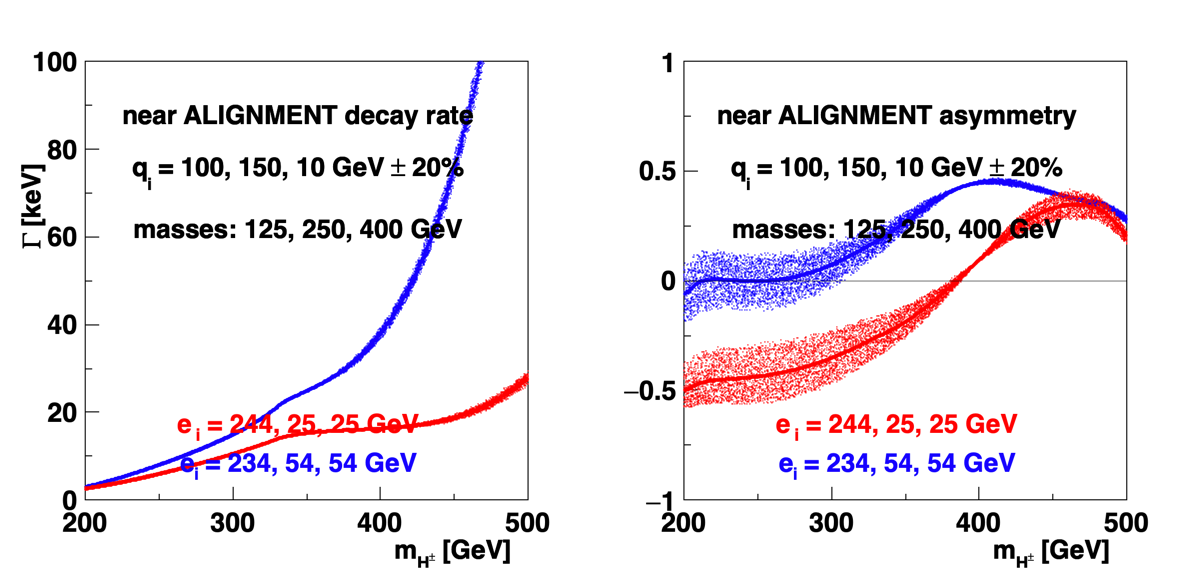}
\includegraphics[scale=0.25]{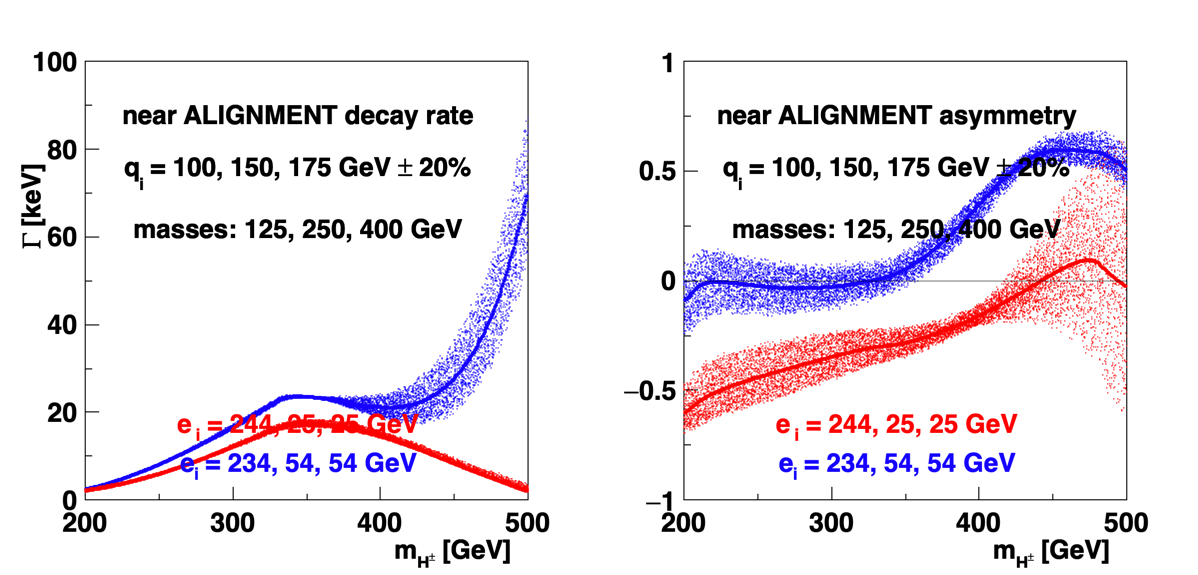}
\end{center}
\vspace*{-4mm}
\caption{Similar to Fig.~\ref{Fig:near-AL+fermions-c}, for $\theta_\text{BF}=0$, $\zeta_t=\pi/4$. Two degrees of alignment are considered. The $q_i$ couplings are taken within 20\% of the  values $\{100, 150, 10\}~\text{GeV}$ (upper row) and $\{100, 150, 175\}~\text{GeV}$ (lower row).}
\label{Fig:near-AL+fermions-q_sccan}
\end{figure}

In Fig.~\ref{Fig:nearAL-rate-asym-r} we illustrated the sensitivity of the decay rate and the charge asymmetry to the couplings $e_2$ and $e_3$, considering a high degree of alignment. Next, in Fig.~\ref{Fig:near-AL+fermions-q_sccan}, we study the dependence on the couplings $q_i$ in the neighbourhood of two points $q_i=\{100, 150, 10\}~\text{GeV}$ and $q_i=\{100, 150, 175\}~\text{GeV}$. Two degrees of alignment are considered, $e_1=0.95v$ (blue) and $e_1=0.99v$ (red). 

The rather large variation of 20\% with respect to the central values considered has only a modest impact on the decay rate and the charge asymmetry in the region of $H^\pm$ masses where the boundedness-from-below constraint is not challenged. However, significant differences arise for higher masses.

{
It may be useful to illustrate the interplay of the bosonic and fermionic sectors further, by displaying in Fig.~\ref{Fig:fig-final} a quantitative comparison of the different contributions, and the resulting interference from the bosonic and fermionic sectors. The adopted parameters coincide with those of the lower panels in Fig.~\ref{Fig:near-AL+fermions-a} and are given in the caption. Left: exact alignment; Right: near-alignment.
Colour coding identifies different contributions. Red: bosonic contribution (on the right, a thin red curve shows the bosonic contribution without fermionic tadpoles; the red curve including fermionic tadpoles is hidden under the purple one); blue: pure fermionic contributions (for better visibility, multiplied by a factor of 10); purple: bosonic plus fermionic contributions (in the full alignment case, we also show, by a thin, purple curve, the result for a reduced fermion coupling, $\tilde\rho_{33}=0.01$); green: including a non-zero value of the inter-sectorial phase $\theta_\text{BF}=\pi/8$ and $\pi/4$, as indicated.

We can make a few additional observations:
(1) In the alignment limit (left panel), the asymmetry is very sensitive to the fermionic contribution, as illustrated by the two purple curves on the left, both very different from the cases of pure bosonic and pure fermionic amplitudes.
(2) Away from alignment (right panel), the asymmetry is much less sensitive to the fermionic amplitude (at this modest level of $\tilde\rho_{33}=0.1$, where the red boson curve is hidden under the purple one).
(3) The sensitivity to the intersectorial phase $\theta_\text{BF}$ is also more pronounced in the limit of alignment.

\begin{figure}[htb]
\begin{center}
\includegraphics[scale=0.27]{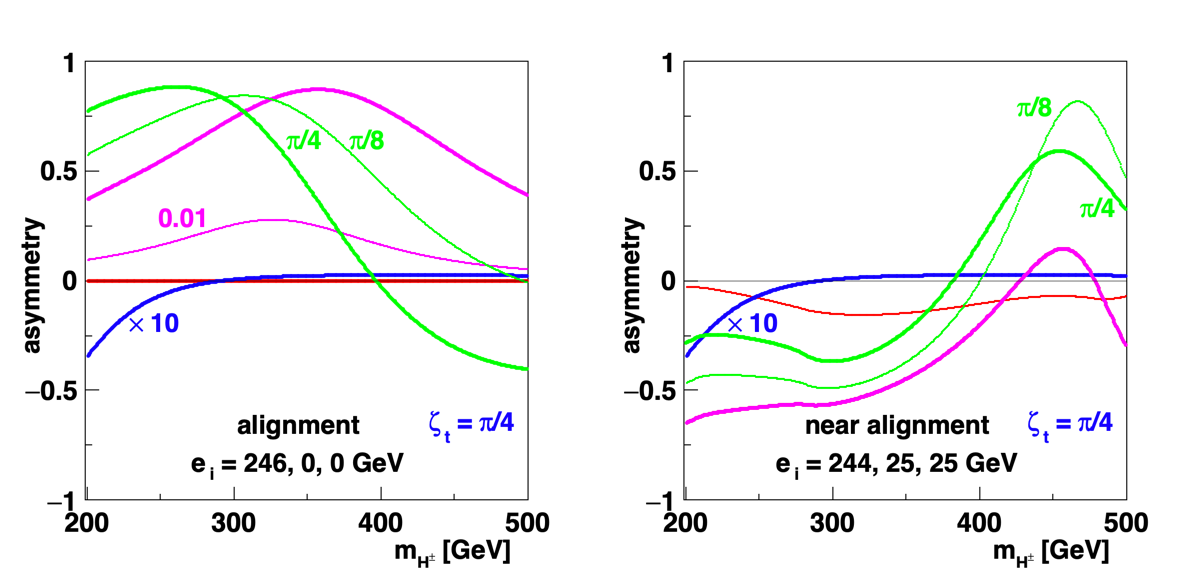}
\end{center}
\vspace*{-4mm}
\caption{{\color{blue}Bosonic sector (red): $m_i=\{125, \,200, \, 400\}~\text{GeV}$;
Left (alignment),  $e_1=v,$ $e_2=e_3=0$; Right: $e_1=0.99v=244~\text{GeV}$, $e_2=e_3=25~\text{GeV}$;
Both: $q_i=\{100, 150, 175\}~\text{GeV}$; \\
Fermionic sector (blue): $m_t=172~\text{GeV}$, $m_b=4.7~\text{GeV}$,
$m_\tau=1.78~\text{GeV}$, $\tilde\rho_{33}^X=0.1\times\exp(i\zeta_X)$,
$\zeta_t=\pi/4$, $\zeta_b=\zeta_\tau=0$;
Purple curves show overall asymmetry (bosonic plus fermionic sectors);
Cross-relation (see section~\ref{sect:Phases}): $\theta_\text{BF}=\pi/8$ , $\pi/4$ (green curves).}}
\label{Fig:fig-final}
\end{figure}
}

\section{Custodial symmetry}
\label{sect:custodial}
Custodial symmetry (CS) is a symmetry of the SM Higgs potential under SU$(2)_L \times$ SU$(2)_R$ that cannot be extended to the gauge and fermionic parts of the Lagrangian. CS is therefore an approximate symmetry of the Lagrangian that is responsible for the electroweak parameter $\rho$ being close to unity. Conditions for CS in the 2HDM were first explored by Pomarol and Vega \cite{Pomarol:1993mu}. In the literature, two different types of custodial symmetries have been studied. However, in \cite{Grzadkowski:2010dj,Haber:2010bw} it has been shown that the two types of CS only differ by a change of basis, and that they are therefore indeed identical despite their different appearance. 

Explicit conditions for the 2HDM potential to be custodial symmetric are given in \cite{Grzadkowski:2010dj,Pilaftsis:2011ed} (in ref.~\cite{Pilaftsis:2011ed}, see symmetry 7 of Table 1). It is worth recalling that these conditions are basis dependent. Basis-independent conditions for CS have been formulated in \cite{Grzadkowski:2010dj,Haber:2010bw}. It is easy to see that the (basis dependent) conditions presented in \cite{Grzadkowski:2010dj,Pilaftsis:2011ed} restrict all parameters of the potential to be real, so that explicit CP violation is not possible. 

It seems to be a standard approach in the literature to make an additional assumption that the vacuum must respect invariance under SU$(2)_\text{diag}$, which constrains the vacuum in such a way that spontaneous CP violation is also not possible, leaving us with a CP conserving Higgs sector.

Here, we will take a different approach. For a custodially symmetric 2HDM potential, we will relax the constraints on the vacuum, leaving open the possibility of spontaneous CP violation. Hence, the Higgs sector may be either CP conserving or violate CP spontaneously.

In \cite{Grzadkowski:2014ada,Grzadkowski:2016szj}, conditions for CP conservation and for spontaneous CP violation have been formulated in terms of physical parameters. There are three cases of CP conservation:\\
\\
{\bf Case A:} $m_1=m_2=m_3$,\\
{\bf Case B:} $m_i=m_j,\, e_iq_j-e_jq_i=0$,\\
{\bf Case C:} $e_k=q_k=0$ for some $k$ (resulting in a CP-odd scalar).\\
\\
If the potential is CP invariant, but none of these three CP conserving cases apply, then we have spontaneous CP violation, characterized by the two constraints:\\
\\
{\bf Case D:} $\displaystyle m_{H^\pm}^2=\frac{v^2\left[
	e_1q_1m_2^2m_3^2+e_2q_2m_1^2m_3^2+e_3q_3m_1^2m_2^2-m_1^2m_2^2m_3^2
	\right]}{2(e_1^2m_2^2m_3^2+e_2^2m_1^2m_3^2+e_3^2m_1^2m_2^2)}$,\\
\phantom{\bf Case D:}	$\displaystyle q=\frac{
		(e_2q_3-e_3q_2)^2m_1^2+(e_3q_1-e_1q_3)^2m_2^2+(e_1q_2-e_2q_1)^2m_3^2+m_1^2m_2^2m_3^2
		}{2(e_1^2m_2^2m_3^2+e_2^2m_1^2m_3^2+e_3^2m_1^2m_2^2)}$.\\
\\
Here, we shall not consider case A nor B, as they have been proven to be RGE unstable \cite{Ferreira:2020ana}. Therefore, a CP conserving Higgs sector must be characterised by case C. 

We shall not perform a detailed analysis here, but briefly sketch how we can arrive at either case C or D, and what are the consequences for the mass of the charged scalars in each case. Let us use the constraints given in symmetry 7 of Table 1 of \cite{Pilaftsis:2011ed}. Solving the stationary point equations for a real vacuum and working out all masses and couplings of such a model, we find that $e_k=q_k=0$ (case C) and $m_{H^\pm}=m_k$ for some $k$. This means that $H_k$ is a CP odd scalar that is mass degenerate with the charged scalars. If instead we solve the stationary point equations for a complex vacuum ($v_{1,2}\neq0,\ \sin\xi\neq0$), we find that the mass of the charged scalars vanishes, $m_{H^\pm}=0$. The constraints of Case D are also found to be satisfied (as they must be), which in turn implies $m_1^2m_2^2m_3^2=e_1q_1m_2^2m_3^2+e_2q_2m_1^2m_3^2+e_3q_3m_1^2m_2^2$, since the charged mass vanishes. The model breaks CP spontaneously, and none of the neutral scalars are CP eigenstates.

It is important to stress that a custodially symmetric 2HDM potential does not necessarily lead to a CP conserving Higgs sector. Likewise, a CP conserving Higgs-sector is not necessarily custodially symmetric. If extended to the fermionic sector, CS would  require the same couplings for up-type as for down-type quarks. Extending CS to the gauge sector also requires the vanishing of hypercharge interactions, leading to $\sin^2\theta_W=0$, in turn implying mass degeneracy between the $W$ and the $Z$ gauge bosons.

If we have the CP conserving version of the CS,
then $e_k=q_k=0,\,M_{H^\pm}=M_k$, and the contribution to the asymmetry from the purely bosonic diagrams interfering among themselves will vanish. There may still be contributions from the purely fermionic diagrams interfering among themselves as well as in the interference between bosonic and fermionic diagrams.

If we have the spontaneously CP violating version of the CS,
then $q$ is given by {\bf Case~D} above, 
$m_{H^\pm}=0$,
$m_1^2m_2^2m_3^2=e_1q_1m_2^2m_3^2+e_2q_2m_1^2m_3^2+e_3q_3m_1^2m_2^2$, and
the contribution to the asymmetry from the purely bosonic diagrams interfering among themselves will not vanish. There will also be contributions from the purely fermionic diagrams interfering among themselves as well as in the interference between bosonic and fermionic diagrams.

{
The relationship betwen CP violation and custodial symmetry may be illustrated by a comparison of parameter spaces. The CP-conserving parameter space is obtained from the CP-violating one by imposing constraints on the latter, i.e.,
\begin{equation}
\Omega(\text{CPC}) \subset \Omega(\text{CPV}),
\end{equation}
where $\Omega$ denotes the parameter space, in particular the number of independent parameters.
Next, we note that custodial symmetry is obtained by imposing further constraints on the CP-conserving parameter space
\cite{Pilaftsis:2011ed}. Then
\begin{equation}
\Omega(\text{CS})\subset \Omega(\text{CPC}) \subset \Omega(\text{CPV}).
\end{equation}
While certain points in $\Omega(\text{CS})$ might be adjacent to some point in $\Omega(\text{CPV})$, there is in general no natural way to pass from $\Omega(\text{CPV})$ to $\Omega(\text{CS})$, without passing through $\Omega(\text{CPC})$, contrary to the impression conveyed in Ref.~\cite{Kanemura:2024ezz}.
}
\section{Summary}
\label{sect:summary}

We have explored the charge asymmetry in the process $H^\pm\to W^\pm Z$ in the 2HDM. Such an asymmetry would arise from underlying CP non-conservation. Even if only one neutral state has a significant coupling to $WW$ and $ZZ$ (let it be $H_1$ represented by the coupling $e_1$) in which case both $e_2$ and $e_3$ would be small, there could still be considerable CP violation caused by the unconstrained couplings $q_2$ and $q_3$ of the states $H_2$ and $H_3$ to $H^+H^-$, see $\Im J_{30}$ in appendix~\ref{sect:CP-2HDM}.
Furthermore, even if the latter were small, 
there could still be a sizable charge asymmetry arising from relative phases among the Yukawa couplings $\tilde\rho^f$ as well as the overall phase between the bosonic and fermionic fields, referred to as $\theta_\text{BF}$ in section~\ref{sect:Phases}.

In the introduction, we mentioned that the CPT theorem dictates that the total decay rates  $\Gamma(H^+)=\Gamma(H^-)$. Thus, there must be some decay channels other than $H^\pm\to W^\pm Z$ that balance out the asymmetry presented in this work. The relevant other decay channels were listed in Eq.~(\ref{eq:decaychannels}). A decay channel will be kinematically forbidden if $m_{H^\pm}<m_a+m_b$, where $m_a$ and $m_b$ are the masses of the two particles that $H^\pm$ decays to. We exploited this fact to do some numerical tests where we (artificially) raised the masses of $H_i$, $t$ and $b$ to see if the symmetry was restored if the masses of the decay products were too large for decay channels involving those particles to be present. We find numerically that with masses such that the decays $H^\pm\to W^\pm H_i$ and $H^\pm \to t\bar b / \bar t b$ are forbidden, the symmetry is restored, $\Gamma(H^+ \to W^+ Z)=\Gamma(H^- \to W^- Z)$, indirectly implying that also $\Gamma(H^+ \to W^+ \gamma)=\Gamma(H^- \to W^- \gamma)$ in this kinematic region. When we lower one mass such that one of the excluded decay channels opens up, the asymmetry reappears. The decay channel $H^\pm \to W^\pm \gamma$ will exhibit similar behavior since it has the same gauge structure as $H^\pm \to W^\pm Z$ and similar loop functions are involved in both these decay channels. This shows that the decay channels listed in Eq.~(\ref{eq:decaychannels}) are all needed in order to restore the symmetry dictated by the CPT theorem. A more technical explanation is that some of the Passarino-Veltman functions resulting from the loop integrals will develop an imaginary part when the mass of the decay products is lower than the mass of the decaying particle.

We confirm qualitative features of the results by Kanemura and Mura \cite{Kanemura:2024ezz}, but also identify some differences: (1) we find a contribution to the charge asymmetry arising from interference among the purely bosonic amplitudes, and (2) we also identify a contribution to the charge asymmetry due to interference among the purely fermionic amplitudes.

The charge asymmetry may well survive in the alignment limit, however  for $m_{2}+m_{3}>m_Z$ (in agreement with experimental results) it would then depend on the fermionic amplitudes.

We conjecture that such a charge asymmetry could be a generic feature of any CP-non\-conserving multi-Higgs-doublet model. The dependence on the underlying physical couplings may be qualitatively similar to what we found for the 2HDM.
The described procedure can be trivially generalised to an NHDM by extending the sums over the fields in the diagrams. However, it should be noted that no $H_i^+ H_j^- Z$ vertices with $i\neq j$ can be generated.

Mathematica files containing the analytic results of our calculation (amplitudes, decay rates and the charge asymmetry) are available at the following GitHub page:
\begin{center}
	\url{https://github.com/omogreid/2HDM-ChargedDecayRates}
\end{center}
These files contain results valid for the general 2HDM with the most general Yukawa matrices, including the possibility for lepton number violation and flavor-changing neutral currents, i.e. more general than the numerical results presented in this paper.
\section*{Acknowledgements}
The authors would like to thank Yushi Mura and Shinya Kanemura for correspondence and clarifications with respect to their earlier work on this topic. WK would like to acknowledge support from the ICTP through the Associates Programme (2022-2027) and CERN Department of Theoretical Physics for their hospitality during her visit in 2025. 
PO is supported in part by the Research Council of Norway.
The work of MNR was partially supported by Funda\c c\~ ao para a Ci\^ encia e a Tecnologia (FCT), Portugal through the projects UIDB/00777/2020, UIDP/00777/2020, UID/00777/2025 \url{https://doi.org/10.54499/UID/00777/2025}, UID/PRR/00777/2025 \url{https://doi.org/10.54499/UID/PRR/00777/2025}, UID/PRR2/00777/2025 \url{https://doi.org/10.54499/UID/PRR2/00777/20250}, CERN/FIS-PAR/0002/2021 and 2024.02004.CERN, which are partially funded through POCTI (FEDER), COMPETE, QREN and the EU. We also thank 
CFTP/ IST/University of Lisbon and the University of Bergen, where collaboration visits took place.

\appendix

\section{CP-violating invariants in the 2HDM}
\label{sect:CP-2HDM}

In the 2HDM, conditions for CP violation/conservation can be expressed in terms of three invariants, here written in terms of masses and couplings \cite{Grzadkowski:2014ada} as
\begin{align} \label{Eq:Im_J1}
\Im J_1&=\frac{1}{v^5}\sum_{i,j,k}\epsilon_{ijk}m_i^2e_ie_kq_j\nonumber\\
&=\frac{1}{v^5}[m_1^2e_1(e_3q_2-e_2q_3)+m_2^2e_2(e_1q_3-e_3q_1)+m_3^2e_3(e_2q_1-e_1q_2)], \\
\Im J_2&=\frac{2e_1 e_2 e_3}{v^9}(m_1^2-m_2^2)(m_2^2-m_3^2)(m_3^2-m_1^2)\nonumber\\
&=\frac{2}{v^9}\sum_{i,j,k}\epsilon_{ijk}e_ie_je_km_i^4m_k^2=\frac{2e_1 e_2 e_3}{v^9}\sum_{i,j,k}\epsilon_{ijk}m_i^4m_k^2, 
 \label{Eq:Im_J2}\\
\Im J_{30}&= \frac{1}{v^5}\sum_{i,j,k}\epsilon_{ijk} q_i m_i^2  e_j q_k. \label{Eq:Im_J3}
\end{align}

The coupling coefficients $e_i$ and $q_i$, which parametrize the couplings of the neutral Higgs bosons to gauge bosons and charged scalars, respectively, are defined in Ref.~\cite{Grzadkowski:2014ada}. We note that all these invariants involve  the antisymmetric $\epsilon_{ijk}$ symbol, involving couplings of {\it all three} neutral states.

\begin{itemize}
\item
If one invariant is non-vanishing, CP is violated in the bosonic sector. \\
\item 
If all invariants vanish, CP is conserved in the bosonic sector. \\
\end{itemize}

\section{Other Feynman rules}
\label{sect:OtherCouplings}
We also use the following Feynman rules (mostly from refs.~\cite{Grzadkowski:2014ada} and \cite{Grzadkowski:2018ohf}):
\begin{eqnarray}
H^+ H^- Z_\mu:&\quad& \frac{ig \cos 2\theta_W}{2\cos\theta_W}(p_--p_+)_\mu,\\
G_0 G^\pm W^\mp_\mu:&\quad& \frac{g}{2}(p_0-p_\pm)_\mu,\\
G^\pm W_\mu^\mp Z_\nu:&\quad& -\frac{ig^2v\sin^2\theta_W}{2\cos\theta_W}g_{\mu\nu},\\
G^+G^-Z_\mu:&\quad&\frac{ig\cos 2\theta_W}{2\cos\theta_W}(p_--p_+)_\mu,\\
H_iH_iH^+G^-:&\quad&
-2i\left[\frac{v^2-e_i^2}{2v^4}(f_1q_1+f_2q_2+f_3q_3)
-\frac{e_if_i}{v^4}m_{H^\pm}^2
+\frac{e_if_i}{v^4}m_i^2 \right. \nonumber\\
&\quad&
\left.-\frac{e_i^2}{2v^6}(e_1f_1m_1^2+e_2f_2m_2^2+e_3f_3m_3^2)\right], \\
H_iH_iH^-G^+:&\quad&
-2i\left[\frac{v^2-e_i^2}{2v^4}(f_1^*q_1+f_2^*q_2+f_3^*q_3)
-\frac{e_if_i^*}{v^4}m_{H^\pm}^2
+\frac{e_if_i^*}{v^4}m_i^2 \right. \nonumber\\
&\quad&
\left.-\frac{e_i^2}{2v^6}(e_1f_1^*m_1^2+e_2f_2^*m_2^2+e_3f_3^*m_3^2)\right], \\
G^0G^0H^+G^-:&\quad&-\frac{i}{v^4}(e_1f_1m_1^2
+e_2f_2m_2^2+e_3f_3m_3^2), \\
G^0G^0H^-G^+:&\quad&-\frac{i}{v^4}(e_1f_1^*m_1^2
+e_2f_2^*m_2^2+e_3f_3^*m_3^2), \\
H^+H^-H^+G^-:&\quad&
-\frac{2i}{v^2}(f_1q_1+f_2q_2+f_3q_3),\\
H^-H^+H^-G^+:&\quad&
-\frac{2i}{v^2}(f_1^*q_1+f_2^*q_2+f_3^*q_3),\\
G^+G^-H^+ G^-:&\quad& -\frac{2i}{v^4}(e_1f_1m_1^2+e_2f_2m_2^2+e_3f_3m_3^2),\\
G^-G^+H^- G^+:&\quad& -\frac{2i}{v^4}(e_1f_1^*m_1^2+e_2f_2^*m_2^2+e_3f_3^*m_3^2),\\
H_iH_iH_i:&\quad&
-6i\left[\frac{v^2-e_i^2}{2v^2}q_i-\frac{(v^2-e_i^2)e_i}{v^4}m_{H^\pm}^2
+\frac{(2v^2-e_i^2)e_i}{2v^4}m_i^2\right] ,\\
H_iH_iH_j:&\quad&
-2i\left[-\frac{e_ie_j}{v^2}q_i+\frac{v^2-e_i^2}{2v^2}q_j
+\frac{(3e_i^2-v^2)e_j}{v^4}m_{H^\pm}^2\right.\nonumber\\
&\quad&\left.\hspace*{2cm}
+\frac{(v^2-e_i^2)e_j}{v^4}m_i^2
-\frac{e_i^2e_j}{2v^4}m_j^2\right],\\
H_i\eta_Z\eta_Z:&\quad&
-\frac{ig^2}{4\cos\theta_W^2}e_i,\\
H_i\eta_W\eta_W:&\quad&
-\frac{ig^2}{4}e_i,
\end{eqnarray}
where the latter are the couplings to Faddeev--Popov ghosts.

\bibliographystyle{JHEP}
\bibliography{ref}

\end{document}